\documentclass[useAMS,usenatbib,epsfig]{mn2e}
\usepackage{graphicx,pdflscape}

\title[The young open cluster IC 1848]{Sejong Open Cluster Survey (SOS). \\
II. IC 1848 Cluster in the H II Region W5 West}
\author[B. Lim et al.]
  {Beomdu Lim,$^1$\thanks{E-mail:bdlim1210@gmail.com}
  Hwankyung Sung,$^1$\thanks{Corresponding author, E-mail:sungh@sejong.ac.kr} 
  Jinyoung S. Kim,$^2$  
  Michael S. Bessell,$^3$ 
 \newauthor and Rivkat Karimov$^4$ \\
  $^1$ Department of Astronomy and Space Science, Sejong University, 209 Neungdong-Ro, Gwangkjin-Gu, Seoul, Republic of Korea\\
  $^2$ Steward Observatory, University of Arizona, 933 N. Cherry Ave. Tucson, AZ 85721-0065, USA\\
  $^3$ RSAA, College of Mathematical and Physical Sciences, The Australian National University, Cotter Road, Weston, ACT 2611, \\ 
  \ \ Australia\\
  $^4$ Ulugh Beg Astronomical Institute, 33 Astronomical Street, Tashkent 700052, Uzbekistan
  }
\date{Released 2013 Xxxxx XX}

\pagerange{\pageref{firstpage}--\pageref{lastpage}} \pubyear{2013}

\def\LaTeX{L\kern-.36em\raise.3ex\hbox{a}\kern-.15em
    T\kern-.1667em\lower.7ex\hbox{E}\kern-.125emX}

\begin{document}

\label{firstpage}

\maketitle

\begin{abstract}
IC 1848 is one of the young open clusters in the giant star forming Cas OB6 association. 
Several interesting aspects relating to star formation processes in giant star forming regions 
attracted us to study the initial mass function (IMF), star formation mode, and properties of pre-main 
sequence stars (PMS). A $UBVI$ and H$\alpha$ photometric study of the young open cluster 
IC 1848 was conducted as part of the ``Sejong Open cluster Survey" (SOS). We have selected 
105 early-type members from photometric diagrams. Their mean reddening is $\langle E(B-V) \rangle = 0.660 
\pm 0.054$ mag. Using the published photometric data with near- and mid-infrared archival data 
we confirmed the normal reddening law ($R_V = 3.1$) toward the cluster (IC 1848). A careful 
zero-age main sequence fitting gives a distance modulus of $V_0-M_V = 11.7 \pm 0.2$ mag, equivalent to $2.2 \pm 0.2$ kpc. 
H$\alpha$ photometry and the list of young stellar objects identified by Koenig et al. permitted us to 
select a large number of PMS stars comprising 196 H$\alpha$ emission stars, 
35 H$\alpha$ emission candidates, 5 Class I, 368 Class II, and 24 transition disk candidates. 
From the Hertzsprung-Russell diagram using stellar evolution models, we estimate an age of 5 Myr 
from several evolved stars and 3 Myr from the PMS stars. The IMF was 
derived from stars with mass larger than $3 M_{\sun}$, and the slope is slightly steeper 
($\Gamma = -1.6 \pm 0.2$) than the Salpeter/Kroupa IMF. Finally, we estimated 
the mass accretion rate of PMS stars with  a UV excess. The mean mass accretion rate is 
about $1.4 \times 10^{-8} M_{\sun} \ \mathrm{yr}^{-1}$ in the mass range of $0.5 M_{\sun}$ to 
$2 M_{\sun}$, whereas intermediate-mass stars ($\ge 2.5 M_{\sun}$) exhibit a much higher 
accretion rate of $\dot{M} > 10^{-6} M_{\sun} \ \mathrm{yr}^{-1}$. 
\end{abstract}


\section{Introduction}

There are three well-known large-scale star forming regions (W3/W4/W5 -- \citealt{W58}) in the Cas 
OB6 association. W5 (also called the IC 1848 H II region) is the eastern part of the molecular cloud 
complex and consists of two components W5 West and W5 East 
\citep{KM03}. There are four known O stars, HD 17505, HD 17520, HD 237019 
and BD +60 586 in W5 West, while one visible O star HD 18326 
is at the center of W5 East. All but HD 237019 are surrounded by many low-mass 
stars in clusters. These O stars are thought to be the ionizing sources of 
the H II region. The simple morphology of each H II region helps us to study the influence 
of strong UV radiation from massive stars on the surrounding materials and the triggering 
mechanisms for star formation \citep{KAG08}. Hence, W5 is an ideal laboratory to 
study star formation processes with feed back from high-mass stars. The initial mass 
function (IMF) of this star forming region is also an interesting scientific issue. In this context, our target of 
interest is the IC 1848 cluster (hereafter IC 1848), also known as OCL 364, situated in the 
center of W5 West, containing the brightest star HD 17505, which is a multiple 
system consisting of an O6.5III, two O7.5V, and an O8.5V stars \citep{HGB06}.   

Many investigators (\citealt{LECT78,VHV79,NTD97} and therein) provided radio maps 
at various frequencies to investigate the large-scale structures and the physical properties of the
W3/W4/W5 complex. Since active star formation takes place in these regions, investigators have 
a continuing interest in the processes involved. \citet{LW78} and 
\citet{TTHRT80} studied star formation in W5 A  (or IC 1848 A), which is a bright-rimmed 
molecular cloud to the eastern end of W5. \citet{WHLJD84} discussed many interesting 
observational aspects associated with the triggered star formation in W5. A census of 
embedded stellar sources in the W3/W4/W5 region have been published by 
\citet{CHS00}. The authors identified 19 clusters, with about half the stars in 5 rich 
clusters. \citet{KM03} identified several young stellar objects (YSOs) using the IRAS point source catalog and found 
that the young stars are mainly distributed around the edge of the H II region. The authors 
argued from a comparison of evolutionary time scales that there is a distinct difference in the 
star formation history of  the ionizing stars and the young stars on the border of expanding 
H II regions. Because of the presence of clumps evaporating away from the ionising sources and the shorter time
scale for star formation under the assumed expanding velocities, they also suggested that 
radiatively driven implosion - star formation triggered by the strong UV radiation from young 
OB stars \citep{EL77} - may be the dominant triggering mechanism in W5.
{\it Spitzer}/IRAC and MIPS imaging data, \citet{KAG08} provided photometric data for more than 
17,000 point sources and YSO classifications. The identified YSOs made it possible to 
study the clustering properties of these YSOs and to discuss plausible 
triggering mechanisms required to explain the distinct generations within W5. The high 
resolution IRAC and MIPS  images revealed several isolated cometary globules and 
elephant trunk structures with polycyclic aromatic hydrocarbon (PAH) emission. Later \citet{KA11} studied 
the disk evolution of intermediate-mass stars. Cometary globules, 
bright-rimmed clouds, and H II region outflows are suggested as evidence of 
triggered star formation, and many studies have investigated the physical properties 
of these objects \citep{LLC97,TWM04,KAKSB08,NTI09}. 

Several photometric studies \citep{S55,JHIMH61,HJI61,BF71,Mo72,LGM01} 
involving IC 1848 have been made in the optical bands. These 
studies provided useful photometric data and fundamental parameters, such as 
reddening, distance, and age for IC 1848. However, most analyses
assumed a normal reddening law ($R_V \sim 3.0$), and 
the multiplicity of early-type stars was not fully taken into account. 
Furthermore, the limiting magnitudes were not deep enough to study the pre-main 
sequence (PMS) stars fainter than $V \sim 16$ mag in the young clusters or associations 
in the Perseus arm. 

Recently, a deep $UBVI_C$ photometric study for IC 1848 W5 
East has been made by \citet{CPO11} with archival data. The authors presented the IMF for stars 
in the mass range from $0.4 M_{\sun}$ to $30 M_{\sun}$, as well as 
their fundamental parameters. However, no deep optical photometric 
study has been made for IC 1848-W5 West. Since the results of \citet{KAG08} 
provided very good membership criteria for the PMS stars with accretion disks, the age 
distribution and mass accretion rates of PMS stars, the star formation history within 
the region, and the IMF down to the $1.0-1.5 M_{\sun}$ regime can be studied. In addition,
the Two Micron All Sky Survey (2MASS) \citep{2mass} and {\it Spitzer}/IRAC data \citep{KAG08} 
make it possible to test carefully the reddening law in a wide wavelength range. 

The previously published distance to IC 1848, derived using various methods, 
is in the range 1.7 to 2.4 kpc. The photometric surveys 
mentioned above provided distances  of 1.7 kpc \citep{S55}, $2.1 \pm 0.3$ kpc 
\citep{CPO11}, $2.2 \pm 0.2$ kpc \citep{JHIMH61,LGM01}, and 2.3 kpc \citep{Mo72,BF71}. Some 
studies have implicitly assumed that the distance to IC 1848 is the same as that to W3 or W4 (or IC 1805) 
(1.9 -- 2.4 kpc). \citet{GG76} independently derived a distance of 2.3 kpc using their Galactic 
rotation model. Early studies have relied on this distance. Recently, \citet{XRZM06} 
derived a distance of $1.95 \pm 0.04$ kpc for a W3OH maser using its VLBI parallax.
 \citet{HGB06} preferred to use a distance of 1.89 kpc to match the 
observed and theoretical stellar radii. Since distance is a critical
parameter in converting observational quantities to reliable physical parameters, 
we need to revisit its determination.

The Sejong Open cluster Survey (SOS) project is dedicated to provide 
homogeneous photometric data down to $V \sim 22$ mag for many open clusters. The 
overview of the project can be found in \citet{over} (Paper 0). The young open cluster 
NGC 2353 was studied as part of this project \citep{LSKI11}. This paper on IC 1848 is 
the third in the series. The observations and comparisons with previous photometry are 
described in Section 2. In Section 3, we discuss the reddening law in the direction of 
 IC 1848 and present fundamental parameters estimated from the 
photometric diagrams. The IMF of IC 1848 and the mass accretion 
rates of PMS stars with UV excesses are presented in Section 4 and 5, respectively. 
Finally, we summarize the results from this study in Section 6.

\begin{figure}
\includegraphics[height=0.45\textwidth]{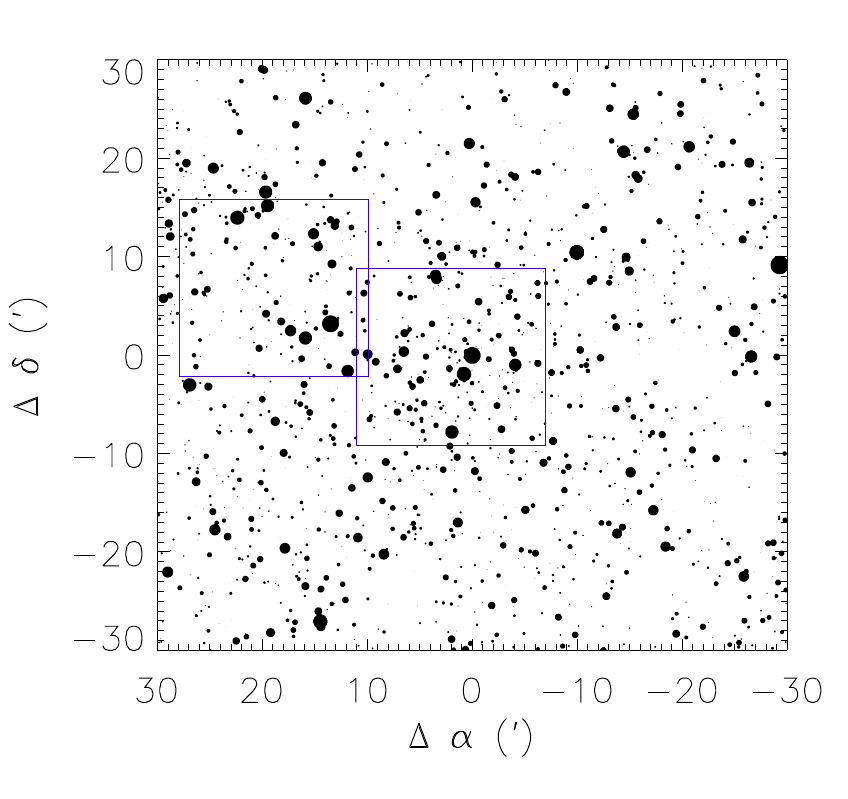}
\caption{Finder chart for IC 1848. The size of the circles are proportional to 
the brightness of individual stars. The position of stars is relative to the brightest O type 
star HD 17505 ($\alpha = 02^h \ 51^m \ 08^{s}.0$ $\delta = +60\degr \ 25' \ 04''$, J2000). 
Squares outline the observed regions.}
\label{fig1}
\end{figure}

\section{OBSERVATION}
The observations of two regions associated with IC 1848 were 
made on 2009 January 19 using the AZT-22 1.5m telescope (f/7.74) at Maidanak 
Astronomical Observatory in Uzbekistan. All imaging data were acquired using 
the Fairchild $4098 \times 4098$ CCD (SNUCam -- \citealt{IKC10}) with standard Bessell $UBVI$ \citep{B90} 
and H$\alpha$ filters. \citet{LSKI08} have described the characteristics 
of the CCD in detail. The mean seeing was better than 1$\farcs0$, and  
sky conditions were good. The observations comprised a total of 20 frames that
were taken in two sets of exposure times for each band -- 5s and 60s in $I$,  5s and 180s in $V$, 
7s and 300s in $B$, 15s and 600s in $U$, and 30s and 600s in H$\alpha$. We 
present the finder chart for the stars brighter than $V = 15$ mag in Figure~\ref{fig1} 
using the Guide Star Catalogue version 2.3 \citep{LLM08}. The photometry for 12 stars 
that were saturated in our data were taken from previous studies. The photometric data and the 
adopted spectral types of the stars are presented in Table~\ref{tab1}. 

\begin{table*}
 \begin{minipage}{150mm}
\caption{Photometric data and spectral types for 12 bright stars}
\begin{tabular}{lccclll}
\hline
ID & $V$ & $B-V$ & $U-B$ & reference & Spectral type & reference \\
\hline
BD +59 556 & 10.06 & 0.36  & -0.12 & \citet{HJI61}    & B9 V       & \citet{HA65}\\
BD +60 584 & 9.87   & 0.28  &           & \citet{HFM00}  &                 &  \\
BD +60 586 & 8.48   & 0.30  & -0.67 & \citet{JH56}    & O7.5 V       &  \citet{HGB06}\\
HD 17505    &  7.06  & 0.40  & -0.64 & \citet{JH56}    & O6.5 III((f)) & \citet{HGB06}\\
HD 17520    &  8.27  & 0.32  & -0.68 & \citet{JH56}    & O9 V       & \citet{CL74}\\
HD 17688    & 10.64 & 0.56  & -0.08 & \citet{B63}      & A7 V        & \citet{F66}\\
HD 237007  &  9.44  & 0.33  & -0.30 & \citet{JH56}    & B0 V        & \citet{MCW55}\\
HD 237011  & 10.05 & 0.30  &           & \citet{HFM00} & B1.5 V       & \citet{R03}\\
HD 237012  & 9.67   & 0.47  &           & \citet{HFM00} &                  &   \\
HD 237015  & 9.44   & 0.24  & -0.43 & \citet{H70}      & B4 D       & \citet{R03}\\
HD 237018  & 9.01   & 0.14  &           & \citet{HFM00} & B9.5 V        & \citet{F66}\\
HD 237019  & 9.73   & 0.47  & -0.53 & \citet{JH56}    & O8 V      & \citet{MCW55}\\
\hline
\label{tab1}
\end{tabular}
\end{minipage}
\end{table*}

As described in \citet{LSKI08}, all pre-processing  to
remove instrumental artifacts was done using the IRAF \footnote{Image Reduction 
and Analysis Facility is developed and distributed by the National Optical 
Astronomy Observatories, which is operated by the Association of Universities 
for Research in Astronomy under operative agreement with the National 
Science Foundation.}/CCRED packages. In order to obtain daily parameters for 
transformation to the standard system, the atmospheric extinction coefficients and 
photometric zero points were obtained on the same night as the cluster observations 
through the observation of many equatorial standard stars \citep{MMLCE91} at different air masses. 
We performed simple aperture photometry for the standard stars with an aperture 
size of $14\farcs0$ and present the coefficients in Table ~\ref{tab2}. We carried out point spread 
function (PSF) photometry for target images using IRAF/DAOPHOT with the small 
fitting radii of 1 FWHM ($\leq 1\farcs0$), and then aperture correction 
was made by applying the differential magnitude between the fitting radii and the 
standard aperture radius ($7\farcs0$). The size of the aperture correction was determined 
from the aperture photometry of relatively bright, isolated stars with a photometric error smaller 
than 0.01 mag for each target image. Finally, the instrumental magnitudes were transformed to the standard magnitude 
and color indices using the transformation equations below \citep{SBCKI08}, 

\begin{equation}
M_{\lambda} = m_{\lambda} - (k_{1\lambda} - k_{2\lambda}C_0)\cdot X + \eta _{\lambda} \cdot C_0 + \alpha _{\lambda} \cdot \hat{UT} + \zeta _{\lambda}
\end{equation}

\noindent where $M_{\lambda}$, $m_{\lambda}$, $k_{1\lambda}$, $k_{2\lambda}$, $\eta _{\lambda}$, 
$C_0$, $X$, $\alpha _{\lambda}$, $\hat{UT}$, and $\zeta _{\lambda}$ are the standard magnitude, 
instrumental magnitude, primary extinction coefficient, secondary extinction coefficient, transformation 
coefficient, relevant color, air mass, time-variation coefficient, time difference relative to midnight, 
and photometric zero point, respectively. A negligible time-variation of $\sim$ 1 mmag per hour in the 
photometric zero points was found in both the $I$ and $V$ bands. We adopted the recently modified transformation 
coefficients ($\eta _{\lambda}$) for SNUCam from \citet{LSBKI09}. From the assumption that the luminosity 
function of all stars in the observed fields has a linear slope across the entire magnitude range we found that 
our photometry is 88\% complete down to $V = 19$ mag, which corresponds to $\sim 1.2 M_{\sun}$. 
The photometric data for 15010 stars from this observation are available in the electronic table (Table~\ref{tab3}) or from 
the authors (B. L. or H. S.).

We checked our photometry against previous studies in order to confirm its consistency. 
Only a few photoelectric photometric data sets for IC 1848 were found in the open cluster 
data base WEBDA\footnote{http://www.univie.ac.at/webda/}. In addition, no photometric study 
with a modern CCD was reported for IC 1848. The photoelectric photometric 
data of \citet{HJI61} were compared with our CCD data for 9-11 common stars. 
The differences in the $V$ magnitude, the $B-V$, and $U-B$ color indices are, respectively, 
$\Delta V = 0.021 \pm 0.020$ mag, $\Delta (B-V) = 0.001 \pm 0.054$ mag, and $\Delta (U-B) 
= 0.008 \pm 0.067$ mag, where $\Delta \equiv$ this - Hoag et al. Although the scatter seems to 
be rather large, our photometric zero points are in good agreement with those of \citet{HJI61}. 
Since the photoelectric data of \citet{HJI61} are well consistent with 
those of \citet{JH56}, we can conclude that our photometry is well tied to the 
Johnson standard $UBV$ system. On the other hand, it is impossible to directly test 
the consistency of our $V-I$ colors due to the absence of photoelectric or CCD 
photometric data. However, we were able to compare our CCD photometric data for the young 
open cluster NGC 1893, which was observed on the same night as IC 1848, with the CCD photometric 
data of other studies and found that our data are well consistent with previous studies 
\citep{MJD95,SPO07}. In the comparisons with the photometric data of other studies 
no evidence for any difference depending on color indices was found. The details of the 
comparisons will be presented in the forthcoming paper (Lim et al. 2013, in preparation). In addition, as 
the ratio of total-to-selective extinction obtained from the color excess ratio of $E(V-I)/E(B-V)$ 
is consistent with that from other color indices (see Figure~\ref{fig6}), our $V-I$ color 
is well tied to the Cousins standard system. 
  
\begin{table}
\begin{minipage}{80mm}
\caption{Extinction coefficients and photometric zero points}
\begin{tabular}{cccc}
\hline
Filter & $k_{1\lambda}$ & $k_{2\lambda}$ & $\zeta_{\lambda}$ (mag)\\
\hline
$I$            & $0.033 \pm 0.014$ &           & $23.604 \pm 0.012$\\
$V$           & $0.134 \pm 0.007$ &           & $24.100 \pm 0.007$\\
$B$           & $0.236 \pm 0.009$ & $0.020 \pm 0.006$ & $24.026 \pm 0.009$\\
$U$           & $0.459 \pm 0.014$ & $0.026 \pm 0.014$ & $22.420 \pm 0.013$\\
H$\alpha$ & $0.081 \pm 0.005$ &           & $20.412 \pm 0.042$ \\
\hline
\label{tab2}
\end{tabular}
\end{minipage}
\end{table}

\section{PHOTOMETRIC DIAGRAMS}
\begin{figure*}
\includegraphics[height=0.45\textwidth]{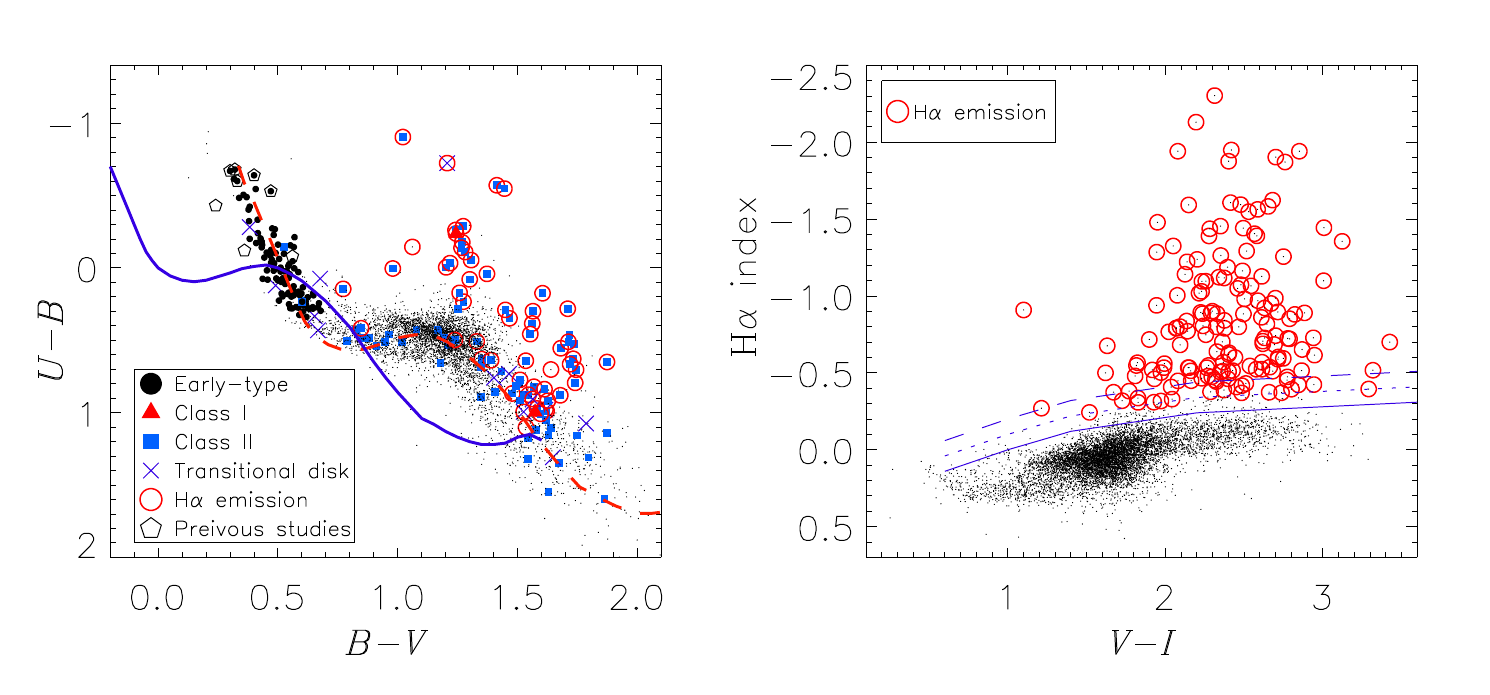}
\caption{Color-color diagrams of IC 1848. In the left panel, bold dots, triangles (red), squares (blue), crosses (blue), 
open circles (red), and pentagons represent early-type members, Class I, Class II, transitional disk candidates, 
H$\alpha$ emission stars, and bright stars obtained from previous studies, respectively. The intrinsic and 
reddened color-color relations are overplotted with a solid and dashed line, respectively. The mean reddening of $\langle E(B-V) \rangle = 
0.66$ mag is adopted for the latter. In the right panel, the solid line represents the empirical photospheric level,
while the dotted and dashed lines are the lower limits of H$\alpha$ emission candidates and H$\alpha$ 
emission stars. From this criteria, 196 H$\alpha$ emission stars and 35 candidates are identified.}
\label{fig2}
\end{figure*}

\begin{figure*}
\includegraphics[height=0.6\textwidth]{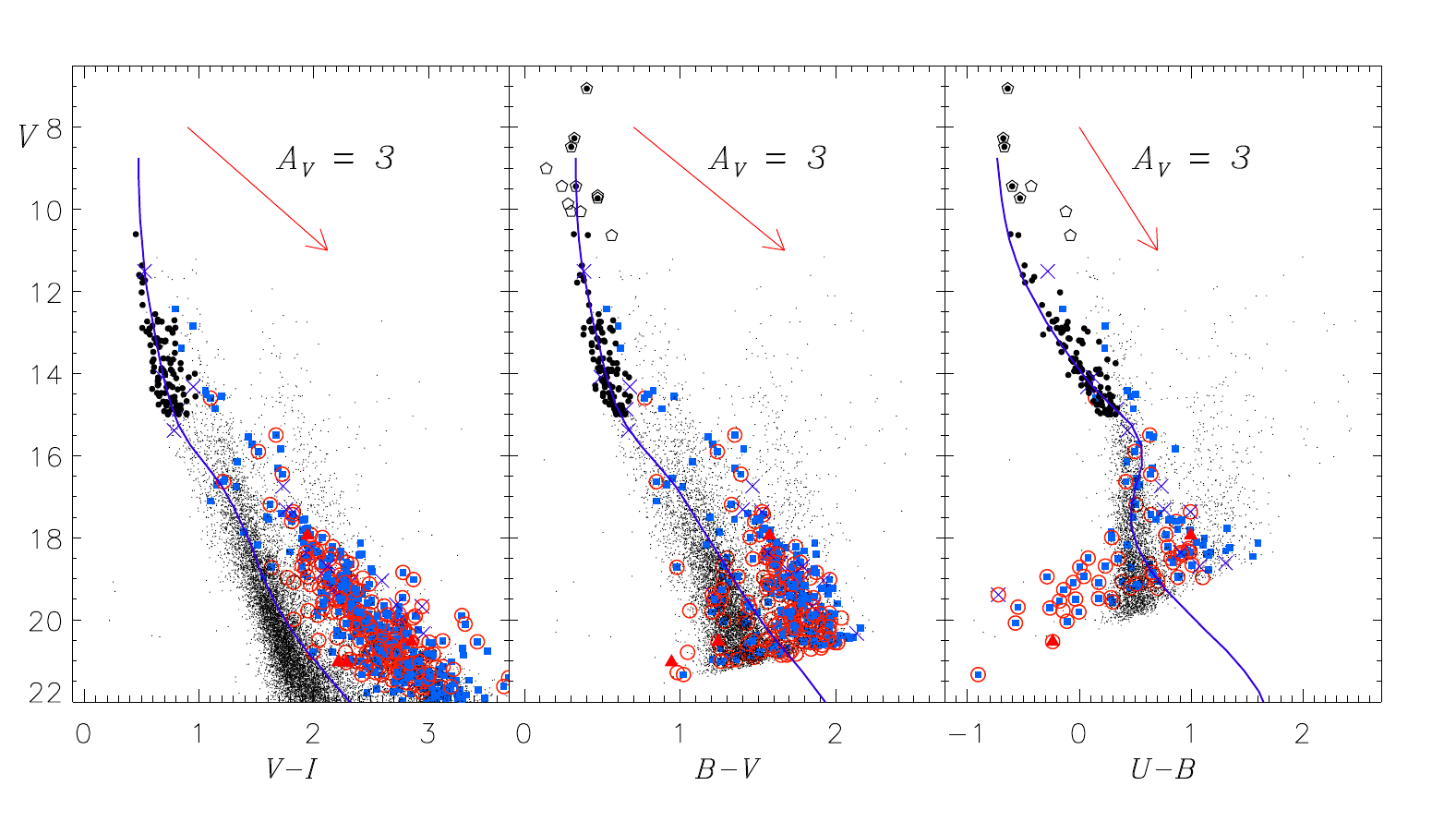}
\caption{Color-magnitude diagrams of IC 1848. Left panel : $V-I$ vs. $V$ diagram. 
Middle panel : $B-V$ vs. $V$ diagram. Right panel : $U-B$ vs. $V$ diagram. The solid lines represent 
the reddened zero-age main sequence relations of \citet{over} shifted by $E(B-V)=0.66$ mag and 
$V_0 - M_V = 11.7$ mag. The arrow denotes a reddening vector corresponding to $A_V = 3$ mag. 
The other symbols are the same as Figure 2. }
\label{fig3}
\end{figure*}

The fundamental parameters of open clusters are very important to study many 
scientific issues, such as the local spiral arm structure of the Galaxy, observational 
test of stellar evolution theory, the stellar IMF, the history of the star formation and chemical 
evolution in the Galactic disk, as mentioned in Paper 0. Such parameters should be 
carefully determined by using homogeneous photometric data. Traditionally, 
fundamental parameters of open clusters are determined from optical 
photometric diagrams. Although near-infrared (NIR) photometry gives much insight into faint or 
obscured objects owing to its less sensitivity to interstellar reddening, optical 
photometry is still a powerful tool to derive many fundamental parameters of 
stars, because it has well-calibrated empirical relations and provides 
relatively higher resolution color indices for massive stars. H$\alpha$ 
photometry \citep{SBL97} and the classifications of YSOs in the W5 region \citep{KAG08} 
provide very useful membership constraints for the PMS stars in IC 1848. In 
this section we present the membership selection criteria, the reddening law, 
and fundamental parameters of IC 1848 from the two-color diagrams (TCD) 
in Figure~\ref{fig2} and the color-magnitude diagrams (CMD) in Figure~\ref{fig3}.

\subsection{Membership Selection}
Early-type members (from O to late-B) can be selected from their photometric properties in the CMDs and 
the $(U-B,B-V)$ TCD. The criteria for the early-type members are (1) $V \leq 15$, $0.2 \leq B-V 
\leq 0.7$, $-1.0 \leq U-B \leq 0.5$ (see left panel of 
Figure~\ref{fig2} and Figure~\ref{fig3}), $E(B-V) \geq 0.5$, and $-1.0 \leq $ 
Johnson's $Q$ $\leq -0.1$, (2) an individual distance modulus between 
$\langle V_0 - M_V \rangle _{\mathrm{cl}} - 0.75 - 2.5\sigma_{V_0 - M_V}$ and 
$\langle V_0 - M_V \rangle _{\mathrm{cl}} + \ 2.5\sigma_{V_0 - M_V}$ 
to take into account the effect of binary members and photometric errors \citep{SB99,KSB10,LSKI11}, 
where $\langle V_0 - M_V \rangle _{\mathrm{cl}}$ and $\sigma_{V_0 - M_V}$ are the mean distance 
modulus and the width of the Gaussian fit of the distance modulus, respectively. The reddening 
of individual members was determined and corrected for (see Section 3.2). The distance to the star was 
estimated using the ZAMS relation from Table 3 in Paper 0. Finally we compared it with 
the average distance and the standard deviation of all members and candidates. In this procedure one star 
(ID 12815) was rejected from the list of members. Another star (ID 6138) was also rejected from the [$E(V-I),E(B-V)$] color excess ratio 
(Figure~\ref{fig6}) due to a large deviation from the global trend. The star may be a foreground 
late-type star. We repeated this procedure until no further stars were rejected. In the end, 105 
early-type stars were selected as members of IC 1848.

Since many PMS stars with an accretion disk exhibit X-ray emission, UV excess, H$\alpha$ emission, 
and an infrared excess, H$\alpha$ photometry is an efficient way to select PMS members 
for young open clusters ($\le 3$ Myr). Indeed, \citet{SBL97} have successfully 
identified PMS members in NGC 2264 using H$\alpha$ photometry. The authors 
and their collaborators have used H$\alpha$ photometry to select PMS 
members in many young open clusters, NGC 6231 
\citep{SBL98,SSB13}, NGC 6530 \citep{SCB00}, NGC 2244 \citep{PS02}, 
NGC 2264 \citep{PSBK00,SBCKI08}, NGC 3603 \citep{SB04}, and 
Trumpler 14 and 16 in the $\eta$ Carina nebula \citep{HSB12}. We found 
196 H$\alpha$ emission stars and 35 candidates using the H$\alpha$ index defined in 
the previous studies \citep{SCB00} and presented the stars (open circle) in the right panel of Figure~\ref{fig2}. 
Stars which exhibit a weak H$\alpha$ emission line are likely missed in a single observation 
because the strength of the H$\alpha$ emission line varies with time. We included 
26 additional stars with emission equivalent widths of H$\alpha$ larger than 5 \AA \ from \citet{KA11}, 
which were not identified as H$\alpha$ emission stars in our photometry.  

Excess emission by dust in the circumstellar disk can be detected at NIR and mid-infrared (MIR) wavelengths. 
The excess due to dust emission is more prominent in the MIR 
than the NIR. But prior to {\it Spitzer}, the Wide-Field 
Infrared Survey (WISE), and {\it Herschel} era, ground-based observations at NIR wavelengths were
restricted to large IR excess stars. As a result, many studies have attempted 
to identify PMS members through strong NIR emission in the ($J-H, H-K$) TCD. Now, owing 
to the vast survey of the W5 region in the MIR \citep{KAG08}, we could easily select the young stars with 
a circumstellar disk. The optical counterparts of the MIR excess emission stars in \citet{KAG08} 
were searched within a searching radius of $1\farcs0$, and we identified 397 YSOs 
(5 Class I, 368 Class II, 24 transition disk candidates). A total 462 PMS members were

\begin{landscape}
\begin{table} {\tiny
\caption{Photometric Data}
\label{tab3}
  \begin{tabular}{rcccccccccccccccccc}
  \hline
  \hline
ID$^1$ & $\alpha_{J2000}$  & $\delta_{J2000}$ & $V$ & $I$ & $V-I$ & $B-V$ & $U-B$ & $H-C$$^2$ &
$\epsilon_V$ & $\epsilon_I$ & $\epsilon_{V-I}$ & $\epsilon_{B-V}$ & $\epsilon_{U-B}$ & $\epsilon_{H-C}$ & N$_{obs}$ & 2MASSID & 
H $\alpha$$^3$ & Sp$^4$\\
  \hline
14941 & 02 54 54.43 & +60 33 01.6 & 21.820  & 19.149  &  2.671  &         &         &         & 0.108   & 0.043   & 0.116   &         &         &         & 1 1 1 0 0 0 &        -         & - &  \\
14942 & 02 54 54.47 & +60 40 27.8 & 19.088  & 17.782  &  1.296  &  0.994  &  0.318  &  0.147  & 0.017   & 0.001   & 0.017   & 0.023   & 0.056   & 0.054   & 1 2 1 1 1 1 &        -         & - &  \\
14943 & 02 54 54.50 & +60 33 38.2 & 20.520  & 18.625  &  1.887  &  1.583  &         & -0.034  & 0.032   & 0.033   & 0.046   & 0.072   &         & 0.102   & 1 1 1 1 0 1 &        -         & - &  \\
14944 & 02 54 54.50 & +60 39 56.9 & 21.789  & 19.122  &  2.667  &         &         &         & 0.092   & 0.046   & 0.103   &         &         &         & 1 1 1 0 0 0 &        -         & - &  \\
14945 & 02 54 54.51 & +60 25 44.9 & 21.871  & 19.314  &  2.557  &         &         &         & 0.110   & 0.058   & 0.124   &         &         &         & 1 1 1 0 0 0 &        -         & - &  \\
14946 & 02 54 54.51 & +60 35 00.0 & 21.333  & 18.909  &  2.419  &  1.021  & -0.904  & -1.949  & 0.059   & 0.029   & 0.066   & 0.103   & 0.142   & 0.075   & 1 1 1 1 1 1 &        -         & H &  \\
14947 & 02 54 54.52 & +60 31 22.2 & 19.802  & 18.202  &  1.586  &  1.210  &  0.489  & -0.027  & 0.024   & 0.009   & 0.026   & 0.038   & 0.082   & 0.068   & 1 2 1 1 1 1 &        -         & - &  \\
14948 & 02 54 54.56 & +60 27 40.6 & 19.663  & 18.043  &  1.606  &  1.165  &  0.394  &  0.053  & 0.023   & 0.003   & 0.023   & 0.035   & 0.097   & 0.060   & 1 2 1 1 1 1 &        -         & - &  \\
14949 & 02 54 54.56 & +60 37 37.6 & 16.727  & 15.435  &  1.287  &  1.158  &  0.662  & -0.019  & 0.007   & 0.015   & 0.017   & 0.007   & 0.014   & 0.017   & 2 2 2 2 1 2 & 02545454+6037376 & - &  \\
14950 & 02 54 54.57 & +60 33 54.0 & 20.510  & 18.792  &  1.705  &  1.344  &         & -0.144  & 0.028   & 0.021   & 0.035   & 0.056   &         & 0.072   & 1 1 1 1 0 1 &        -         & - &  \\
14951 & 02 54 54.58 & +60 30 03.0 & 16.264  & 15.055  &  1.200  &  0.916  &  0.395  &  0.159  & 0.002   & 0.001   & 0.002   & 0.003   & 0.003   & 0.007   & 2 2 2 2 2 2 & 02545459+6030029 & - &  \\
14952 & 02 54 54.61 & +60 23 46.0 & 20.390  & 18.553  &  1.821  &  1.354  &         &  0.066  & 0.043   & 0.043   & 0.061   & 0.080   &         & 0.144   & 1 1 1 1 0 1 &        -         & - &  \\
14953 & 02 54 54.61 & +60 39 26.0 & 20.104  & 18.541  &  1.550  &  1.155  &         &  0.043  & 0.030   & 0.038   & 0.048   & 0.041   &         & 0.100   & 1 1 1 1 0 1 &        -         & - &  \\
14954 & 02 54 54.66 & +60 27 10.0 & 21.281  & 19.344  &  1.937  &         &         &         & 0.057   & 0.044   & 0.072   &         &         &         & 1 1 1 0 0 0 &        -         & - &  \\
14955 & 02 54 54.66 & +60 39 33.2 & 19.562  & 18.122  &  1.429  &  1.138  &  0.311  &  0.107  & 0.015   & 0.004   & 0.016   & 0.029   & 0.065   & 0.064   & 1 2 1 1 1 1 &        -         & - &  \\
14956 & 02 54 54.68 & +60 26 41.0 & 21.290  & 18.269  &  3.021  &         &         & -0.316  & 0.070   & 0.019   & 0.072   &         &         & 0.111   & 1 2 1 0 0 1 & 02545466+6026409 & - &  \\
14957 & 02 54 54.68 & +60 26 50.0 & 20.026  & 18.183  &  1.827  &  1.343  &         & -0.083  & 0.027   & 0.023   & 0.035   & 0.048   &         & 0.076   & 1 1 1 1 0 1 &        -         & - &  \\
14958 & 02 54 54.71 & +60 35 18.5 & 20.586  & 19.003  &  1.570  &  1.179  &         &         & 0.034   & 0.042   & 0.054   & 0.073   &         &         & 1 1 1 1 0 0 &        -         & - &  \\
14959 & 02 54 54.72 & +60 25 15.3 & 20.986  & 19.271  &  1.714  &         &         &         & 0.047   & 0.067   & 0.082   &         &         &         & 1 1 1 0 0 0 &        -         & - &  \\
14960 & 02 54 54.75 & +60 39 16.6 & 19.986  & 18.361  &  1.615  &  1.388  &         & -0.147  & 0.025   & 0.042   & 0.049   & 0.043   &         & 0.080   & 1 1 1 1 0 1 &        -         & - &  \\
14961 & 02 54 54.77 & +60 24 46.8 & 21.401  & 19.193  &  2.208  &         &         &         & 0.080   & 0.054   & 0.097   &         &         &         & 1 1 1 0 0 0 &        -         & - &  \\
14962 & 02 54 54.83 & +60 23 25.4 & 19.485  & 17.504  &  1.972  &  1.599  &         & -0.068  & 0.025   & 0.042   & 0.049   & 0.046   &         & 0.084   & 1 2 1 1 0 1 & 02545478+6023253 & - &  \\
14963 & 02 54 54.87 & +60 31 30.0 & 20.260  & 18.669  &  1.583  &  1.375  &         &  0.136  & 0.029   & 0.023   & 0.037   & 0.054   &         & 0.091   & 1 1 1 1 0 1 &        -         & - &  \\
14964 & 02 54 54.88 & +60 32 32.9 & 17.459  & 16.242  &  1.208  &  0.947  &  0.261  &  0.110  & 0.001   & 0.003   & 0.003   & 0.006   & 0.013   & 0.005   & 2 2 2 2 1 2 & 02545487+6032330 & - &  \\
14965 & 02 54 54.89 & +60 24 18.7 & 20.259  & 18.320  &  1.927  &  1.467  &         &  0.167  & 0.034   & 0.036   & 0.050   & 0.060   &         & 0.114   & 1 1 1 1 0 1 & 02545493+6024182 & - &  \\
14966 & 02 54 54.89 & +60 37 15.3 & 19.560  & 18.101  &  1.447  &  1.075  &  0.438  & -0.003  & 0.018   & 0.002   & 0.018   & 0.030   & 0.078   & 0.069   & 1 2 1 1 1 1 &        -         & - &  \\
14967 & 02 54 54.93 & +60 26 36.8 & 21.570  & 19.357  &  2.213  &         &         &         & 0.074   & 0.047   & 0.088   &         &         &         & 1 1 1 0 0 0 &        -         & - &  \\
14968 & 02 54 54.93 & +60 36 08.3 & 21.135  & 19.113  &  2.022  &         &         &         & 0.048   & 0.053   & 0.072   &         &         &         & 1 1 1 0 0 0 &        -         & - &  \\
14969 & 02 54 54.95 & +60 34 02.7 & 21.471  & 19.511  &  1.960  &         &         &         & 0.065   & 0.054   & 0.085   &         &         &         & 1 1 1 0 0 0 &        -         & - &  \\
14970 & 02 54 54.97 & +60 34 35.6 & 21.595  & 19.833  &  1.762  &         &         &         & 0.067   & 0.082   & 0.106   &         &         &         & 1 1 1 0 0 0 &        -         & - &  \\
14971 & 02 54 54.98 & +60 27 39.9 & 20.853  & 18.842  &  2.009  &  1.454  &         &  0.124  & 0.049   & 0.035   & 0.060   & 0.106   &         & 0.117   & 1 1 1 1 0 1 &        -         & - &  \\
14972 & 02 54 54.99 & +60 37 45.4 & 15.276  & 14.075  &  1.195  &  1.045  &  0.478  &  0.051  & 0.008   & 0.017   & 0.019   & 0.008   & 0.003   & 0.019   & 2 2 2 2 2 2 & 02545498+6037454 & - &  \\
14973 & 02 54 55.04 & +60 24 44.7 & 21.713  & 19.558  &  2.156  &         &         &         & 0.113   & 0.072   & 0.134   &         &         &         & 1 1 1 0 0 0 &        -         & - &  \\
14974 & 02 54 55.11 & +60 37 04.1 & 21.434  & 19.870  &  1.564  &         &         &         & 0.066   & 0.071   & 0.097   &         &         &         & 1 1 1 0 0 0 &        -         & - &  \\
14975 & 02 54 55.13 & +60 29 01.7 & 20.780  & 18.986  &  1.783  &  1.507  &         &  0.037  & 0.045   & 0.029   & 0.054   & 0.085   &         & 0.120   & 1 1 1 1 0 1 &        -         & - &  \\
14976 & 02 54 55.15 & +60 26 11.1 & 20.804  & 18.855  &  1.949  &         &         & -0.079  & 0.046   & 0.033   & 0.057   &         &         & 0.122   & 1 1 1 0 0 1 &        -         & - &  \\
14977 & 02 54 55.16 & +60 26 15.1 & 20.760  & 18.893  &  1.853  &  1.397  &         &         & 0.042   & 0.053   & 0.068   & 0.080   &         &         & 1 1 1 1 0 0 &        -         & - &  \\
14978 & 02 54 55.18 & +60 35 31.8 & 21.857  & 19.427  &  2.430  &         &         &         & 0.102   & 0.068   & 0.123   &         &         &         & 1 1 1 0 0 0 &        -         & - &  \\
14979 & 02 54 55.22 & +60 23 05.0 & 19.978  & 18.181  &  1.784  &  1.414  &         &  0.079  & 0.032   & 0.038   & 0.050   & 0.058   &         & 0.117   & 1 1 1 1 0 1 &        -         & - &  \\
14980 & 02 54 55.22 & +60 31 44.3 &         & 19.540  &         &         &         &         &         & 0.063   &         &         &         &         & 0 1 0 0 0 0 & 02545522+6031443 & - &  \\
14981 & 02 54 55.22 & +60 40 05.1 & 21.551  & 18.649  &  2.902  &         &         & -1.392  & 0.088   & 0.043   & 0.098   &         &         & 0.112   & 1 1 1 0 0 1 & 02545519+6040051 & - &  \\
14982 & 02 54 55.23 & +60 25 34.1 & 19.176  & 17.586  &  1.578  &  1.193  &  0.307  &  0.074  & 0.020   & 0.017   & 0.026   & 0.033   & 0.066   & 0.058   & 1 2 1 1 1 1 &        -         & - &  \\
14983 & 02 54 55.23 & +60 27 15.2 & 20.342  & 18.611  &  1.718  &  1.334  &         & -0.034  & 0.037   & 0.026   & 0.045   & 0.072   &         & 0.094   & 1 1 1 1 0 1 &        -         & - &  \\
14984 & 02 54 55.25 & +60 33 23.7 & 17.478  & 16.143  &  1.326  &  1.035  &  0.382  &  0.127  & 0.002   & 0.003   & 0.003   & 0.007   & 0.015   & 0.005   & 2 2 2 1 1 2 & 02545527+6033237 & - &  \\
14985 & 02 54 55.25 & +60 40 00.5 & 20.193  & 18.297  &  1.883  &  1.442  &         & -0.171  & 0.026   & 0.037   & 0.045   & 0.048   &         & 0.069   & 1 1 1 1 0 1 &        -         & - &  \\
14986 & 02 54 55.28 & +60 38 42.1 & 17.320  & 15.974  &  1.334  &  0.988  &  0.483  &  0.093  & 0.002   & 0.042   & 0.042   & 0.009   & 0.018   & 0.042   & 2 2 2 2 1 2 & 02545526+6038420 & - &  \\
14987 & 02 54 55.29 & +60 25 00.7 & 20.842  & 18.941  &  1.899  &  1.488  &         &         & 0.059   & 0.073   & 0.094   & 0.112   &         &         & 1 1 1 1 0 0 &        -         & - &  \\
14988 & 02 54 55.29 & +60 34 43.6 & 20.481  & 18.792  &  1.682  &  1.499  &         & -0.072  & 0.036   & 0.030   & 0.047   & 0.080   &         & 0.099   & 1 1 1 1 0 1 &        -         & - &  \\
14989 & 02 54 55.31 & +60 34 31.8 & 17.953  & 16.599  &  1.344  &  1.003  &  0.380  &  0.151  & 0.006   & 0.015   & 0.016   & 0.011   & 0.020   & 0.025   & 2 2 2 1 1 1 & 02545530+6034317 & - &  \\
14990 & 02 54 55.33 & +60 37 24.0 & 20.629  & 18.956  &  1.662  &  1.361  &         & -0.010  & 0.030   & 0.033   & 0.045   & 0.079   &         & 0.133   & 1 1 1 1 0 1 &        -         & - &  \\
14991 & 02 54 55.35 & +60 33 59.5 & 21.502  & 18.803  &  2.699  &         &         &         & 0.080   & 0.035   & 0.087   &         &         &         & 1 1 1 0 0 0 &        -         & - &  \\
14992 & 02 54 55.37 & +60 34 58.4 & 18.915  & 17.508  &  1.396  &  1.083  &  0.320  &  0.064  & 0.011   & 0.029   & 0.031   & 0.019   & 0.040   & 0.046   & 1 2 1 1 1 1 & 02545535+6034583 & - &  \\
14993 & 02 54 55.46 & +60 33 24.3 & 21.086  & 18.306  &  2.780  &         &         &  0.291  & 0.065   & 0.009   & 0.066   &         &         & 0.114   & 1 2 1 0 0 1 &        -         & - &  \\
14994 & 02 54 55.48 & +60 37 45.6 & 20.792  & 17.944  &  2.848  &         &         &         & 0.046   & 0.027   & 0.053   &         &         &         & 1 2 1 0 0 0 &        -         & - &  \\
14995 & 02 54 55.55 & +60 34 06.8 & 19.123  & 17.497  &  1.616  &  1.351  &  0.965  & -0.095  & 0.014   & 0.021   & 0.025   & 0.025   & 0.148   & 0.039   & 1 2 1 1 1 1 & 02545561+6034065 & - &  \\
14996 & 02 54 55.58 & +60 40 30.2 & 21.191  & 19.150  &  2.042  &         &         &         & 0.071   & 0.055   & 0.090   &         &         &         & 1 1 1 0 0 0 &        -         & - &  \\
14997 & 02 54 55.60 & +60 40 50.4 & 18.871  & 17.106  &  1.755  &  1.494  &  1.027  & -0.164  & 0.015   & 0.015   & 0.021   & 0.029   & 0.095   & 0.068   & 1 2 1 1 1 1 & 02545556+6040503 & - &  \\
14998 & 02 54 55.62 & +60 33 27.1 & 16.102  &         &         &  0.882  &         &         & 0.012   &         &         & 0.018   &         &         & 1 0 0 1 0 0 & 02545562+6033271 & - &  \\
   -1 & 02 51 07.97 & +60 25 03.9 &  7.060  &         &         &  0.400  & -0.640  &         &         &         &         &         &         &         & 1 0 0 1 1 0 & 02510801+6025037 & - & O6.5III((f)) \\
   -2 & 02 51 14.46 & +60 23 09.8 &  8.270  &         &         &  0.320  & -0.680  &         &         &         &         &         &         &         & 1 0 0 1 1 0 & 02511443+6023099 & - & O9V \\
   -3 & 02 50 34.90 & +60 24 06.7 &  9.440  &         &         &  0.330  & -0.600  &         &         &         &         &         &         &         & 1 0 0 1 1 0 & 02503490+6024067 & - & B0V \\
   -4 & 02 52 00.89 & +60 25 27.2 & 10.060  &         &         &  0.360  & -0.120  &         &         &         &         &         &         &         & 1 0 0 1 1 0 & 02520088+6025271 & - & B9V \\
   -5 & 02 51 35.71 & +60 32 51.4 & 10.050  &         &         &  0.300  &         &         &         &         &         &         &         &         & 1 0 0 1 0 0 & 02513571+6032513 & - & B1.5V \\
   -6 & 02 51 36.48 & +60 33 11.7 &  9.670  &         &         &  0.470  &         &         &         &         &         &         &         &         & 1 0 0 1 0 0 & 02513647+6033116 & - &  \\
   -7 & 02 54 10.66 & +60 39 03.5 &  8.480  &         &         &  0.300  & -0.670  &         &         &         &         &         &         &         & 1 0 0 1 1 0 & 02541067+6039036 & - & O7.5V \\
   -8 & 02 53 11.37 & +60 37 25.5 &  9.870  &         &         &  0.280  &         &         &         &         &         &         &         &         & 1 0 0 1 0 0 & 02531138+6037254 & - &  \\
   -9 & 02 53 16.94 & +60 26 50.2 &  9.010  &         &         &  0.140  &         &         &         &         &         &         &         &         & 1 0 0 1 0 0 & 02531691+6026501 & - & B9.5 V \\
  -10 & 02 53 28.47 & +60 27 35.0 &  9.730  &         &         &  0.470  & -0.530  &         &         &         &         &         &         &         & 1 0 0 1 1 0 & 02532845+6027348 & - & O8.0V \\
  -11 & 02 52 44.14 & +60 23 29.0 &  9.440  &         &         &  0.240  & -0.430  &         &         &         &         &         &         &         & 1 0 0 1 1 0 & 02524414+6023289 & - & B4D \\
  -12 & 02 52 57.60 & +60 28 16.8 & 10.640  &         &         &  0.560  & -0.080  &         &         &         &         &         &         &         & 1 0 0 1 1 0 & 02525759+6028167 & - & A7V \\
\hline
\end{tabular}
\begin{tabular}{@{}l@{}}
$^1$ The negative numbered ID represents the data from Johnson \& Hiltner (1956); Hoag et al. (1961); Bigay (1963); Haug (1970); H$\o$g et al. (2000) \\
$^2$ $H-C$ color index represents the H$\alpha$ index [$\equiv (V+I)/2$] \\
$^3$ H: H $\alpha$ emission stars; h: H $\alpha$ emission star candidates \\
$^4$ Spectral type -- Morgan et al. (1955); Hoag \& Applequist (1965); Fehrenbach (1966); Conti \& Leep (1974); Reed (2003); Hillwig et al. (2006) \\ 
\end{tabular}}
\end{table}
\end{landscape}

\noindent  identified in our photometry.

As seen in the ($V, V-I$) CMD of Figure~\ref{fig3} or 8, our membership selection 
constrains the PMS locus well. However, current membership are certainly incomplete, 
because the membership criteria of PMS stars mentioned above are biased toward stars with an 
active accretion disk. However, it is worth comparing the detection efficiency of low-mass PMS stars 
between H$\alpha$ photometry and MIR data for stars brighter than $V = 19$ (our photometric 
completeness is about or better than 88\%). A total of 94 stars were classified as low-mass PMS 
members either from MIR excess emission or H$\alpha$ photometry. Among them, 32 
stars were detected from both MIR and H$\alpha$ photometry. Interestingly, none of 5 bright 
transitional disk candidates ($V \leq 16$) shows H$\alpha$ emission. However, the H$\alpha$ 
detection fraction of faint transitional disk candidates ($3/8 = 38 \pm 22$\% to $V = 16$--19 mag) 
is very similar to that of Class II objects ($26/77 = 34\pm7$\%). This fact implies that 
the faint transitional disks have similar characteristics to Class II objects, i.e. they are 
probably pre-transition disk objects \citep{SSB09}. The reason for the absence of H$\alpha$ emission among 
bright PMS stars with a transitional disk may be related either to the absence of a
hot inner disk or to weak H$\alpha$ emission with H$\alpha$ emission 
equivalent widths smaller than 10 \AA \citep{SBCKI08}, or both. The detection 
efficiency of transitional disk candidates ($V \leq 19$ mag) from H$\alpha$ 
photometry may be less than 23 \% with respect to MIR excess emission. In this work, MIR excess 
appears to be a more efficient way to find PMS members than H$\alpha$ photometry. 

However, the detection efficiency from MIR and H$\alpha$ photometry is not as 
high as X-ray observation and strongly depends on the age of a cluster because 
the excess emission is prominent in young PMS stars with a circumstellar disk and 
accretion activity. In addition, we know that weak-line T-Tauri stars (WTTS) are 
unlikely to be selected as PMS members from MIR or H$\alpha$ photometry 
because most of the material in the disk of such stars has dissipated. Indeed, in NGC 2264, 
many WTTS without H$\alpha$ emission or a MIR excess  were identified as PMS by X-ray observations
\citep{SSB09}, and their photometric properties were found to be similar to those of MS stars 
\citep{FMS99}. Almost complete lists of PMS members in NGC 2264 and NGC 6231 have 
now been identified down to their limiting magnitude using X-ray sources 
from the {\it Chandra} data archive \citep{SBCKI08} or from XMM-Newton 
observation \citep{SRNGV06}. Up to now, no extensive X-ray observations 
for the IC 1848 region exist, so we note that a large number of WTTS or low-mass PMSs 
with weak accretion activity are likely missed in our membership selection.

\begin{figure}
\includegraphics[height=0.4\textwidth]{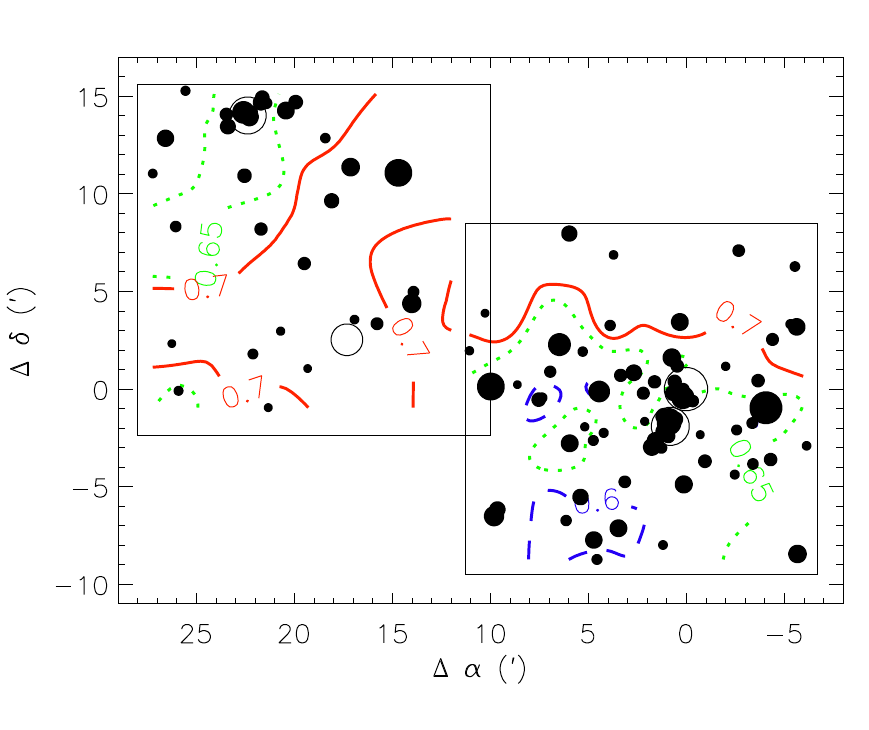}
\caption{Reddening map of IC 1848. Squares outline the observed regions. 
The size of the filled circles is proportional to the brightness of individual early-type stars. Reddening 
contour lines are shown for $E(B-V) =$ 0.60 (dashed line), 0.65 (dotted line), and 0.70 mag (solid line). Large open circles
indicate the O stars. }
\label{fig4}
\end{figure}

\subsection{The Reddening}
The interstellar reddening toward young open clusters is, in general, derived 
by comparing the observed color indices of the early-type members with the intrinsic 
($U-B, B-V$) diagram. Using the intrinsic color-color relation from Table 1 in Paper 0 we  determined the individual 
reddening $E(B-V)$ of early-type members of IC 1848. The reddening slope was simply assumed to be 
$E(U-B)/E(B-V) = 0.72$ because the dependence on $E(B-V)$ is negligible for less reddened 
stars [$E(B-V) < 1$ mag]. The mean reddening value was estimated to be $\langle E(B-V) \rangle = 0.660 \pm 
0.054$ (s.d.) mag and displayed in the left panel of Figure~\ref{fig2} as a dashed line. This value 
is in good agreement with that of previous studies, e.g., $E(B-V)$ = 0.55 -- 0.70 mag -- \citep{JHIMH61}, 
0.66 mag -- \citep{BF71}, 0.72 mag -- \citep{Mo72}, and 0.60 mag -- \citep{LGM01}. 

\begin{figure*}
\includegraphics[height=0.45\textwidth]{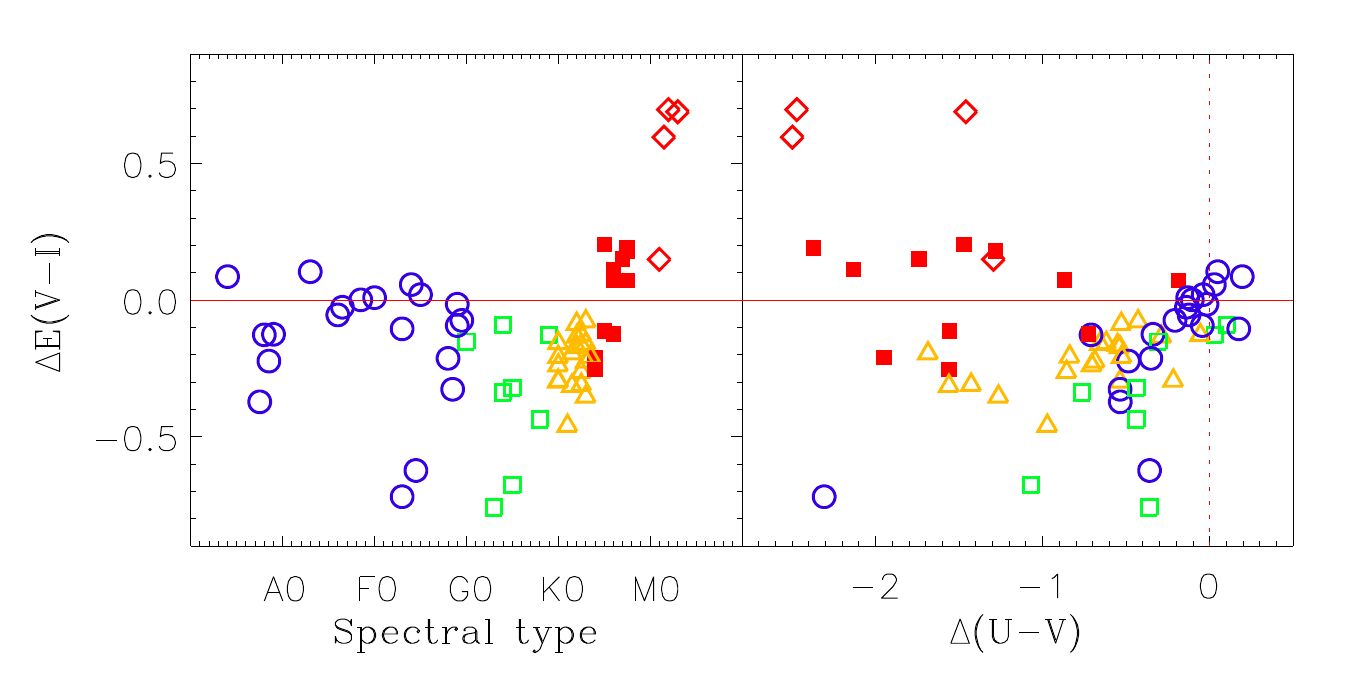}
\caption{Differences between the reddening adopted from the reddening map and reddening
derived from the spectral-type intrinsic $V-I$ color relation plotted against spectral type and UV excess.   
$\Delta$ means the reddening from Figure~\ref{fig4} minus the spectroscopic 
reddening. Circles, open squares, triangles, filled squares, and diamonds represent 
B0-G0, G1-K0, K1-K4, K5-K9, and M type stars, respectively. $\Delta (U-V)$ means UV excess derived from 
this work. The stars with negative $\Delta (U-V)$ values are UV excess stars. See the main text for details.}
\label{fig5}
\end{figure*}

From the spread of the early-type members in the $(U-B, B-V)$ diagram, one can 
see that there is a non-negligible amount of differential reddening across the observed 
region. In order to investigate the spatial variation of the reddening we constructed the reddening 
map of the observed regions using the position and reddening of individual early-type members. 
The observed field was divided into 10,000 rectangular areas, each of which has the size of 
$0\farcm33 \times 0\farcm23$ in right ascension and declination, respectively. 
The weighted-mean reddening of each point is calculated from the
reddening of 105 early-type members, where the weight is exponentially decreasing with the 
distance from individual early-type members. We present the reddening map of 
IC 1848 in Figure~\ref{fig4}. The massive molecular cloud 
W5 NW in W5 West lies in the northern part of the cluster (see Figure 1 of \citealt{WHLJD84} 
and \citealt{KAG08}). O type stars, HD 17505, HD 17520, and HD 237007, in W5 West 
are responsible for the ionization of the molecular clouds \citep{KM03}, and bright rims lie along 
the ionization front \citep{WHLJD84}. These could be evidence that the stellar wind and 
UV radiation from three O type stars are compressing the molecular cloud. On the other hand, 
the reddening map of the southern part is rather transparent compared to the northern part, and 
one could expect that the material in the southern part might have been rapidly 
swept away. Thus, the large scale variation of reddening shows the systematic 
difference between the northern and southern part of the cluster. Another O star BD +60 
586 ($\Delta \alpha \sim 22\farcm5, \Delta \delta \sim 14\farcm0$) is located in the valley of 
reddening between two molecular clouds W5 NE and W5 NW. The structure in the reddening 
map may reflect the interaction between the O stars and the surrounding materials. A similar 
feature was found in NGC 6231 \citep{SSB13}. From this, we deduce that the strong stellar 
wind from O stars can sweep away the surrounding medium. In contrast, \citet{LCS13} found 
the highest reddening value to be in the center of the starburst cluster Westerlund 1 and 
interpreted this as resulting from dust formed from the ejected material from the many 
Wolf-Rayet stars. These observations indicate that within IC 1848, dust destruction may be 
more dominant than dust production. However, unlike the other O type stars, HD 237019 
($\Delta \alpha \sim 17\farcm3, \Delta \delta \sim 2\farcm5$) 
lies in a highly reddened region. According to a formation scenario of W5 West suggested 
by \citet{KAG08} the star could be ejected from the cluster by gravitational interaction 
between massive stars in the cluster center. If this scenario is true, the star may not have had 
enough time to interact with the surrounding material.

To check the reliability of the reddening corrections for PMS stars in our analysis, 
we compared the reddening $E(V-I)$ from Figure~\ref{fig4} with that from the spectral 
type. Recently \citet{KA11} published the spectral types of 389 stars in the W5 region. Among them,
63 stars with $V-I$, $B-V$, and $U-B$ color indices in our data, were used in this comparison. 
The spectroscopic reddening was obtained by comparing 
the observed $V-I$ color with the intrinsic color inferred from a given 
spectral type (Paper 0). One can expect that the mean difference would be close to zero with 
a random scatter, and that spectroscopic reddening could be higher for few stars due 
to their internal extinction. We present the comparison in the left panel of Figure~\ref{fig5}, 
where negative values mean that spectroscopic reddening is larger than that from our 
reddening map. For B -- early-F and K5 -- late-K type stars, the reddening shows reasonably good agreement. However, 
the spectroscopic reddening is 0.2 mag higher than ours for late-F -- K4 type 
stars, and systematically lower for M type stars.

In order to find the main causes of the differences we checked our intrinsic relations in 
Paper 0, which is based on the ($B-V$, $V-I$) relation by \citet{C78} with the spectral type - color 
relation of \citet{BB88}. The difference is less than 0.02 mag, and therefore it is 
unlikely to be caused by the spectral type -- ($V-I$) relation. We also compared the 
reddening map from optical photometry with that from the 2MASS data. With $E(V-H)$ and $E(V-K_S)$ 
of the early-type members (see next section) the NIR reddening map was constructed through 
the same procedure as the optical reddening map. The spatial variation of the $E(H-K_S)$ reddening 
is compatible with that of the optical photometry. In addition, we also checked the spatial distribution of the
G -- M type stars which showed a systematic difference in reddening derived from the two different methods. If our 
reddening map exclusively provides small/large value of reddening toward any 
specific direction, this method of reddening correction may be inappropriate for 
the PMS members. However, the late-type PMS stars were distributed uniformly, thus the 
application of the reddening map may have nothing to do with the discrepancy. 

\begin{figure*}
\includegraphics[height=0.48\textwidth]{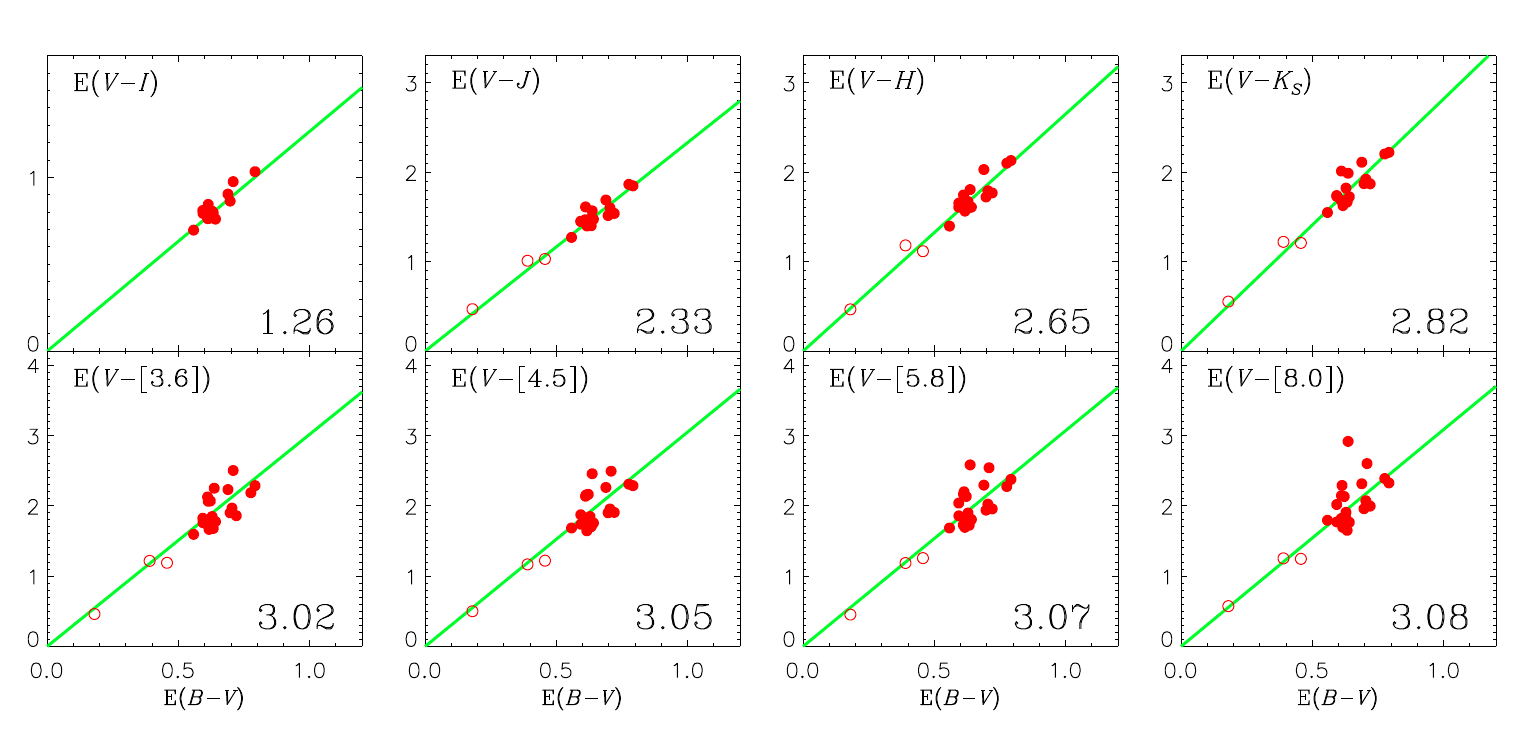}
\caption{Color excess ratios obtained from the early-type stars. Open circles and 
filled circles represent foreground stars and the members of the cluster ($U-B \leq -0.2$), respectively. The solid line 
corresponds to the normal reddening law ($R_V = 3.1$). We present each color excess 
ratio corresponding to $R_V = 3.1$ in the bottom of each panel. The color excess ratios inferred from the 
optical to mid-infrared data consistently shows the standard reddening law.}
\label{fig6}
\end{figure*}

The other possibility is that the spectral types of the stars were classified as earlier 
types due to veiling filling in some of the lines. \citet{KA11} described their 
classification schemes without any discussion on the veiling effect. Since 
the software SPTclass \citep{HCB04} provides the spectral type based on the equivalent width of 
the spectral lines of O8--M6 type MS stars (see their webpage\footnote{http://dept.astro.lsa.umich.edu/$\sim$hernandj/SPTclass/sptclass.html}), veiling effect could make the spectral classification of PMS stars tricky. The classification schemes in SPTclass 
are optimized for Herbig Ae/Be, late F-- mid K, and mid K -- M type stars, respectively. The 
classification for late F -- mid K type stars includes 11 spectral indices in the relatively narrow spectral 
region (from 4200 \AA \ to 6200 \AA), and thus it can be affected by veiling. Furthermore there are fewer 
number of spectral indices than for Herbig Ae/Be and later type stars. Hence, the spectral 
classification of PMS stars could be misled by the presence of non-photospheric contributions. 
According to the spectral indices for F, G, and K type stars, e.g. Ca I $\lambda4226$, Fe I $\lambda$4458, and Fe I $\lambda$5329, 
the equivalent width of the lines increases with spectral type. If optical spectra of PMS stars with veiling 
exhibit decreased line depths, the equivalent widths of the spectral indices will appear 
to be that of earlier types. Therefore veiling would shift the spectral classification to earlier 
types. If the speculation is correct, the stars should exhibit a UV excess. We plot UV excess, $\Delta (U-V)$, 
derived in Section 5, with respect to the discrepancy in the reddening $E(V-I)$ in the right panel of Figure~\ref{fig5}. 
The stars possibly classified as earlier spectral type are expected to be located in the lower left side of 
the panel. Indeed, they (F0 -- K4 type PMS stars - circle, open square, and triangle) locate 
in the expected region, while stars without a UV excess are clustered at $\Delta E(V-I) = 0$ 
and $\Delta (U-V) = 0$.

\begin{figure*}
\includegraphics[height=0.25\textwidth]{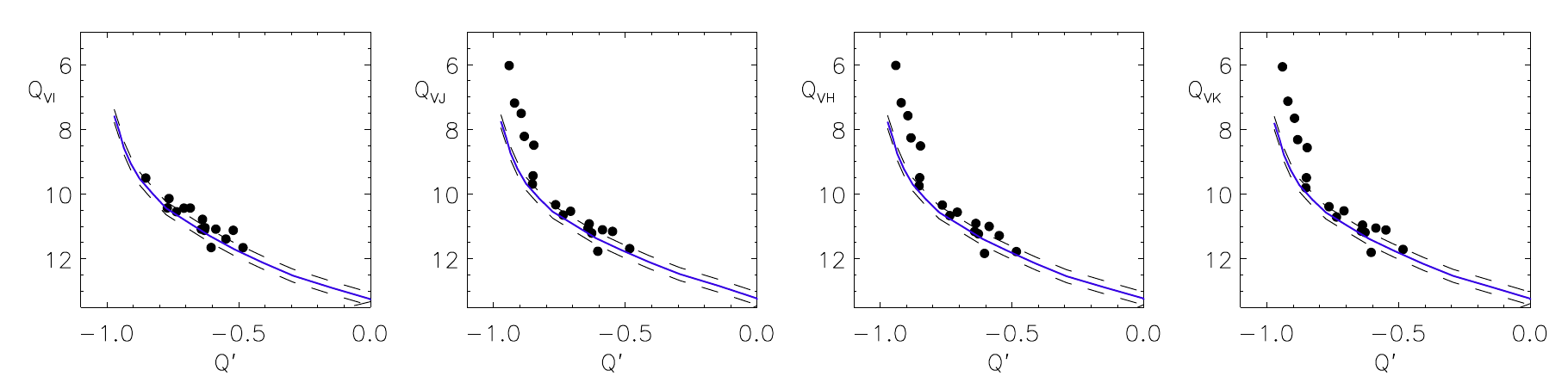}
\caption{Determination of the distance to IC 1848. Dots represent the early-type members ($U-B \leq -0.2$). 
In order to determine the distance to the cluster, the zero-age main sequence relations of \citet{over} are 
used after shifting by $11.7 \pm 0.2$ mag, respectively. The solid line (blue) corresponds to the adopted 
value of 2.2 kpc, and the dashed lines are ZAMS relations shifted by the uncertainties.}
\label{fig7}
\end{figure*}

However, K5 -- M type stars show a systematic and opposite trend. As the strength of TiO and CaH bands are
used for the classification of mid-K to M type stars, the spectral classification should be correct.
In addition, these red bands would be little affected by non-photospheric contributions. As shown in
the right panel of Figure~\ref{fig5}, K5 -- K9  type stars (filled square) are close 
to zero in $\Delta E(V-I)$, although most of them exhibit a UV excess. However, 3 (ID 10938, 13480, and 14932)
out of 4 M type stars with $V-I$ and $U-V$ color indices show a small spectroscopic reddening
$E(V-I)_{\mathrm{sp}}$ of $\sim 0.2$ mag, which is an unrealistic value for IC 1848. We also checked the position
of the stars in Figure~\ref{fig8} and found them to lie in the PMS locus. These facts imply that they have
too-late a spectral type. According to \citet{KA11} those stars are identified as Class II objects with strong
H$\alpha$ emission, and our H$\alpha$ photometry also tags them as H$\alpha$ emission stars.
They must be very young stars of IC 1848, and therefore we attribute this discrepancy to a gravity
effect on the spectra of young PMS stars. Many studies \citep{LLR97,LBRH98,BHSM98,WGRS99}
remark that the spectral features of young PMS stars between 6000 \AA \ and 9000 \AA \
resemble those of giant stars and we note that the inclusion in the spectral classification methodology of
a Na I $\lambda8183/8195$ index, that is very gravity sensitive, would help discriminate between
giant- and dwarf-like M spectra. Given the weaker TiO bands of MS stars compared to giant stars,
it is possible that the stronger bands of a PMS star can lead to them being assigned a later dwarf spectral 
type. 

The same trend as in Figure~\ref{fig5} can also be found in the published data ($A_V$ and spectral type) of 
\citet{KA11}. We investigated their $A_V$ values with respect to the spectral types. The $A_V$ values 
of G -- K type stars appear to be higher than those of other spectral types. The mean $A_V$ of B -- F0 type 
stars is about 1.5 mag, but the $A_V$ values of G5 type stars have a range of 2.0-2.6 mag. 
The differences between two groups with different spectral types in $E(V-I)$ is 
about 0.21-0.45 mag. These values are in good agreement with those found in Figure~\ref{fig5}. In 
addition, the $A_V$ values of K -- M type stars decrease with the subclasses. The feature is the same as that 
found in Figure~\ref{fig5}. These facts imply that the discrepancy found in Figure~\ref{fig5} originates 
from the published data of \citet{KA11}. We therefore conclude that the application of our 
reddening map is more reliable than the reddening correction inferred from previously published spectral 
types in the current state and does not cause serious problems, except for a few PMS stars with nearly edge-on disks.

\subsection{The Reddening Law}
Dust evolution in star forming regions is one of the more interesting issues in astronomy, since 
the grain size is directly related to the reddening law toward an object of interest. 
Many studies \citep{RL85,GV89,LSKI11} have confirmed 
that the extinction law in the Solar neighbourhood, the Galactic center, or in moderately young open 
clusters and associations is nearly the normal law (a total-to-selective extinction ratio of 
$R_V \sim 3.1$). However the value is significantly larger for several extremely young
 star forming regions (see Table 3 in \citealt{Gr10}). For IC 1848 several previous photometric 
studies \citep{JHIMH61,BF71,Mo72} have assumed a normal reddening law. Recently, \citet{CPO11} 
found different $R_V$ in four different regions within W5 East. Since the ratio of 
total-to-selective extinction is a useful tool to study the state of dust evolution 
in star forming regions, and since the value is critical in the canonical determination of distance, 
we tested the reddening law in IC 1848 using color excess ratios. 

Using the NIR 2MASS data and MIR {\it Spitzer} data of \citet{KAG08} it is possible
to determine the reddening law toward IC 1848 in a consistent way from the optical 
to the MIR photometry. To calculate the color excess ratios $ E(V - \lambda) / E(B-V)$, 
observed color indices were compared with the intrinsic colors for the 2MASS bands, from Table 2 in Paper 0,
and the {\it Spitzer} IRAC bands (Sung et al. 2013, in preparation). In this analysis, we used 21 of
the most probable normal MS stars among 105 early-type members with $U-B$ color index 
smaller than -0.2 mag, to exclude the young active stars which could potentially affect the 
color excess ratios. In addition, three foreground stars were also used to investigate the foreground 
reddening law. The color excess ratios obtained from such stars are plotted in Figure~\ref{fig6}. 

We confirmed that the color excess ratios of foreground stars (open circle) and early-type 
members (dot) of IC 1848 are well-consistent with the solid line corresponding to 
the normal reddening law ($R_V = 3.1$). Unlike the optical band and the NIR band, there is a slightly 
large scatter in the MIR bands. There are two possible sources of scatter. We carried out 
PSF photometry for the point sources in four sets of the long-exposed {\it Spitzer} 
images (Program ID: 20300, PI: Allen, Lori) for the W5 region. The photometric data were 
compared with those of \citet{KAG08} and the Glimpse 360 catalogue where 
3.6 and 4.5 \micron \ bands are available. The photometric data are consistent with each 
other within 0.05 mag in the zero points, except for the 4.5 \micron \ band 
($\Delta [4.5] = 0.07-0.08$ mag). We also found a scatter of 0.03-0.07 mag among 
MIR data, and therefore some stars can show a deviation of up to 0.1 mag. In addition, 
the inherent uncertainty of MIR photometric data arising from the nonlinearity of the MIR detector 
and the difficulty in the flux calibration is, in general, up to 10\%. The photometric 
error could be also increased for some stars lying in crowded fields or PAH emission nebulae. 
On the other hand, it is a well-known fact that IR excess 
emission arises from a circumstellar disk or surrounding material of young stars. Thus, 
the scatter in the MIR bands may be partly due to uncertainties in the MIR photometry and 
partly due to circumstellar dust emission of the stars. Allowing for the excess emission, 
the ratios of total-to-selective extinction ($R_V$), which we obtained here, are consistent 
for all wavelengths, and therefore the reddening law toward IC 1848 is normal. It implies that 
the size distribution of dust in IC 1848 is very similar to that of the general diffuse interstellar 
medium. \citet{CPO11} found a normal reddening law toward the cluster around HD 18326 
in the W5 East H II region, while the $R_V$ of bright-rimmed 
clouds shows various values from 2.8 to 3.5. The authors argued that the dust grains in the 
regions have inhomogeneously evolved. Given their results, the strength of UV radiation 
from O type stars in the clusters affects the evolution of dust grains in the natal cloud.

\subsection{Distance to IC 1848}

ZAMS fitting is the most important method to estimate the distance to open clusters. 
In order to obtain a reliable distance from all CMDs, reddening should be corrected for
properly. As mentioned above, the total extinction in the $V$ band for individual 
early-type members was determined from $A_V = R_V \times E(B-V)$. The mean value was $\langle A_V \rangle = 
2.05 \pm 0.17$ (s.d.) mag. After correcting for the reddening and extinction, ZAMS 
fitting to the de-reddened CMDs was performed. Since multiplicity and evolution 
effects make stars brighter, ZAMS fitting to the lower ridge of the MS band 
is recommended, allowing for the photometric error. Several relatively faint early-type 
members (mid -- late B type) appears to be bluer than the ZAMS color indices. \citet{SBLL02} 
have reported the correlation between UV excess and X-ray luminosity of MS stars in the intermediate-age 
open cluster NGC 2516 (age $\sim$160 Myr). According to them, the higher the X-ray luminosity, the 
bluer the $U-B$ and $B-V$ color indices. Due to the absence of X-ray observations for 
IC 1848, we could not obtain the X-ray luminosity of the blue early-type members. 
In the case of the young open cluster NGC 1893, the late B type MS stars with bluer color indices 
were found to be X-ray emitters (Lim et al. 2013, in preparation). If this property is 
universal, high X-ray luminosity is responsible for the bluer color indices. If the ZAMS 
relations were fitted to such blue MS stars, the distance would be 
overestimated. Thus, the most probable normal MS members ($U-B \leq -0.2$), 
as done in the previous section, were used in the ZAMS fitting procedure. From the ZAMS fitting to 
all reddening-corrected CMDs, the distance modulus of IC 1848 was estimated to be 
$V_0 - M_V = 11.7$ mag, which is equivalent to 2.2 kpc. In Figure~\ref{fig7}, we also 
obtained the same result from four reddening-independent indices, as shown in Paper 0 
(Equation 9 -- 12). The uncertainty in the distance modulus may be 0.2 mag from 
Figure~\ref{fig7}.

The distance derivations of previous investigators ranged from 1.7 kpc to 2.4 kpc. 
\citet{S55} obtained a distance of 1.7 kpc from their spectral types and photometric 
data for O stars. However, the author did not discuss the treatment of extinction 
correction nor considered binarity. The distance to the cluster from other photometric 
studies \citep{JHIMH61,BF71,Mo72,GS92,LGM01,CPO11} clusters around  
$2.2 \pm 0.1$ kpc, assuming the normal reddening law. There are two independent 
measurements. \citet{GG76} constructed new Galactic rotation curves derived from the
distances of exciting stars and H$\alpha$ radial velocities for the northern and 
southern hemispheres, respectively. The authors obtained a distance of 2.3 kpc for 
IC 1848 from a $V_{\mathrm{LSR}}$ of $-36.6 \ \mathrm{km} \ \mathrm{s}^{-1}$. This value is consistent 
with that of the photometric studies. However, the adopted galactocentric distance (10 kpc) of the Sun 
is rather larger than the currently accepted value. 

On the other hand, \citet{XRZM06} measured a 
trigonometric parallax for 12-GHz methanol masers in the W3OH region and obtained a distance 
of $1.95\pm0.04$ kpc. Given the little annual variation of methanol masers, the result seems to be 
reliable. Although there is an implicit assumption that the distances to W3/W4/W5 are the 
same, no careful discussion of the assumption has been made. \citet{SHH08} and \citet{SSM13} 
have measured the trigonometric parallax for the young open cluster NGC 281 and two 
H$_2$O maser sources, IRAS 00259+5625 and IRAS 00420+5530, using VLBI Exploration 
of Radio Astrometry. According to their studies, the three H$_2$O maser sources 
($d = 2.82^{+0.26} _{-0.22}$ kpc, $2.43^{+1.02} _{-0.56}$ kpc, and $2.17^{+0.05} _{-0.05}$ kpc) 
are along the line of sight, sequentially (see Figure 5 in \citealt{SSM13}). From these measurements 
they could estimate the size of the expanding ring of molecular clouds, which was found by 
\citet{MBD02,MBD03}, where the diameter along the line of sight is larger than 650 pc. 
Given these studies, the difference between the distances estimated from the photometric 
methods and the trigonometric parallax for the W3OH region ($\sim$ 250 pc) can be interpreted as 
the extent along the line of sight, and therefore the W3OH region is likely in front of the W3/W4/W5 
complex. Since our result supports the estimates from previous 
photometric studies for IC 1848, we use the distance modulus (11.7 mag) obtained here. 
In Figure~\ref{fig3} we present the CMDs with a ZAMS obtained by adopting the mean reddening and distance.

\begin{figure}
\includegraphics[height=0.41\textwidth]{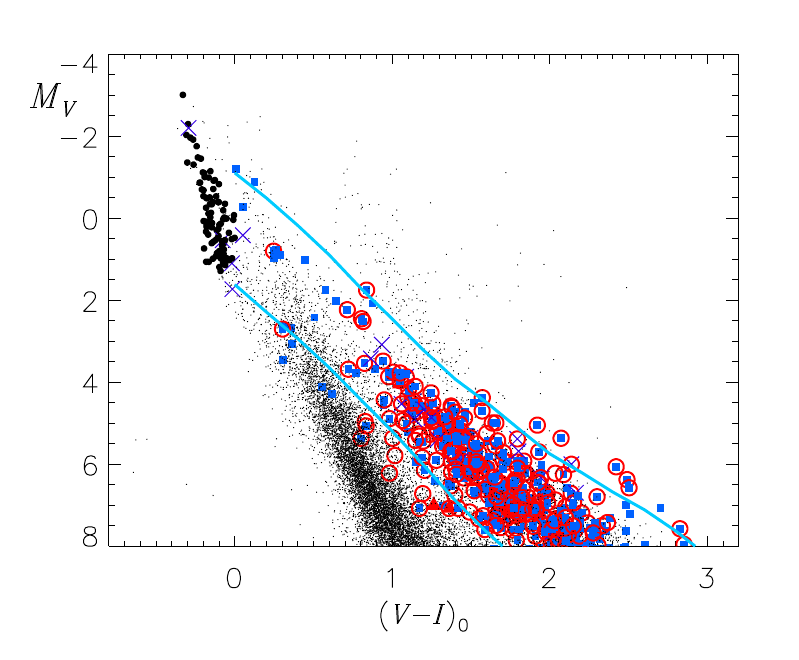}
\caption{$M_V$ vs. $(V-I)_0$ color-magnitude diagram. The reddening of early 
type stars is estimated from the $U-B$ vs $B-V$ color-color 
diagram. The reddening map was used to correct the reddening of low-mass PMS members. 
The pre-main sequence locus is defined by the two solid lines which enclose the distribution of most PMS members 
in the diagram. The other symbols are 
the same as Figure 2.}
\label{fig8}
\end{figure}

\subsection{The Hertzsprung-Russell Diagram and Age of IC 1848}
We directly corrected for the reddening of individual early-type members, while 
the reddening map (Figure~\ref{fig4}) was used for the reddening correction of the other 
members. The dereddened $(V, V-I)$ CMD is presented in Figure~\ref{fig8}. In order 
to constrain the upper limit of the IMF (see next section), the locus of PMS stars was 
defined as shown in the figure (solid lines). 

\begin{figure}
\includegraphics[height=0.42\textwidth]{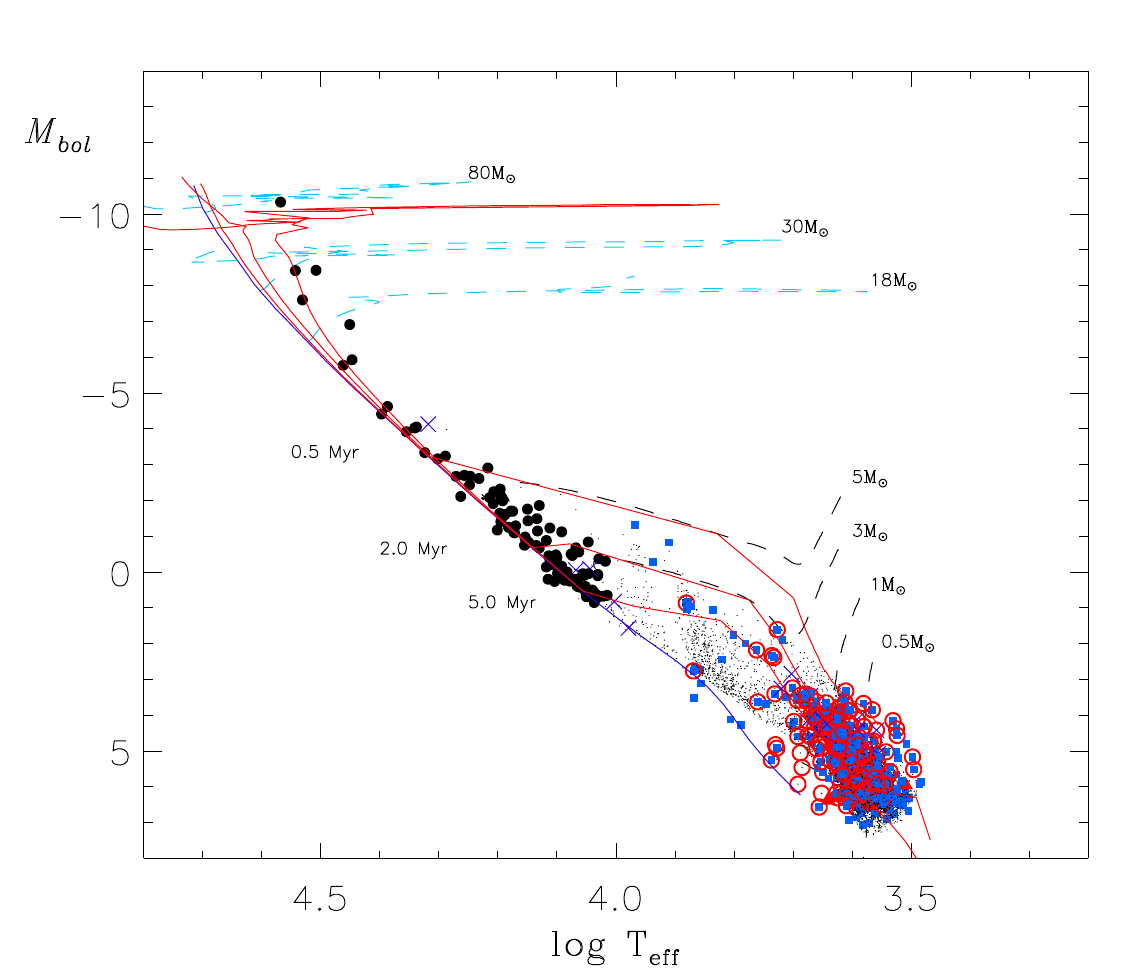}
\caption{The Hertzsprung-Russell diagram of IC 1848. Several isochrones (0, 0.5, 2.0, 
and 5.0 Myr) are superimposed on the diagram with several evolutionary tracks 
\citep{rot12,SDF00}. The other 
symbols are the same as Figure 2.}
\label{fig9}
\end{figure}

We describe the calibrations required for the construction of the Hertzsprung-Russell diagram (HRD). For stars 
earlier than O9, the spectral type -- effective temperature relation from Table 5 in Paper 0 
was used to estimate the effective temperature. We applied the relation between 
$(V-I)_0$ and effective temperature of \citet{BCP98} to the PMS members with $(V-I)_0 \leq 1.4$ 
and another relation \citep{B95} to the remaining PMS members to avoid 
the effects of the UV excess, arising from accretion activities, on the $U-B$ and/or $B-V$ colors. 
For the other stars we averaged the temperature derived from the spectral type -- effective temperature 
relation and the color -- effective temperature relations with an appropriate weight. 
Finally the bolometric corrections were obtained from the relations in Paper 0. 
We present the HRD of IC 1848 in Figure~\ref{fig9}. 

Several evolutionary tracks (dashed lines) with an initial mass and isochrones (solid lines) 
were over plotted in the figure. The stellar evolution models of \citet{rot12}, which take 
into account the effect of stellar rotation on the evolution of stars for a given mass, 
were used for post-MS and MS stars, while we used the models of \citet{SDF00} for 
PMS stars. The isochrones (0.5, 2.0, and 5.0 Myr) have been constructed from the two 
models. The stars with masses larger than $18 M_{\sun}$ seem to be moving away from the
MS stage, while intermediate-mass stars $(\sim 3 M_{\sun})$ seem to be approaching 
the ZAMS. The isochrone with an age of 5 Myr well predicts the position of the 
most massive stars in the HRD. However, according to \citet{HGB06}, HD 17505, HD 17520, 
and BD+60 586 are binaries or multiple systems, and therefore 
it is probable that the turn-off age may be overestimated. We have conducted a
deconvolution of the brightness of HD 17505 by assuming that the 
multiple system consists of an O6.5 III, two O7.5V stars, and an O8.5 V \citep{HGB06}. As a result, 
the HD 17505 system (the O6.5III component deconvolved from the others) should be dimmed 
by $\sim 0.58$ mag from its current position, and is then well consistent with the 
5 Myr isochrone. Although we adopted 
a luminosity class III for the brightest component of HD 17505, it is worth 
noting that the genuine luminosity may be very close to Ib or Iab supergiant because the 
deconvolved absolute magnitude is -6.29 mag (see Table 4. in Paper 0). 

\begin{figure}
\includegraphics[height=0.56\textwidth]{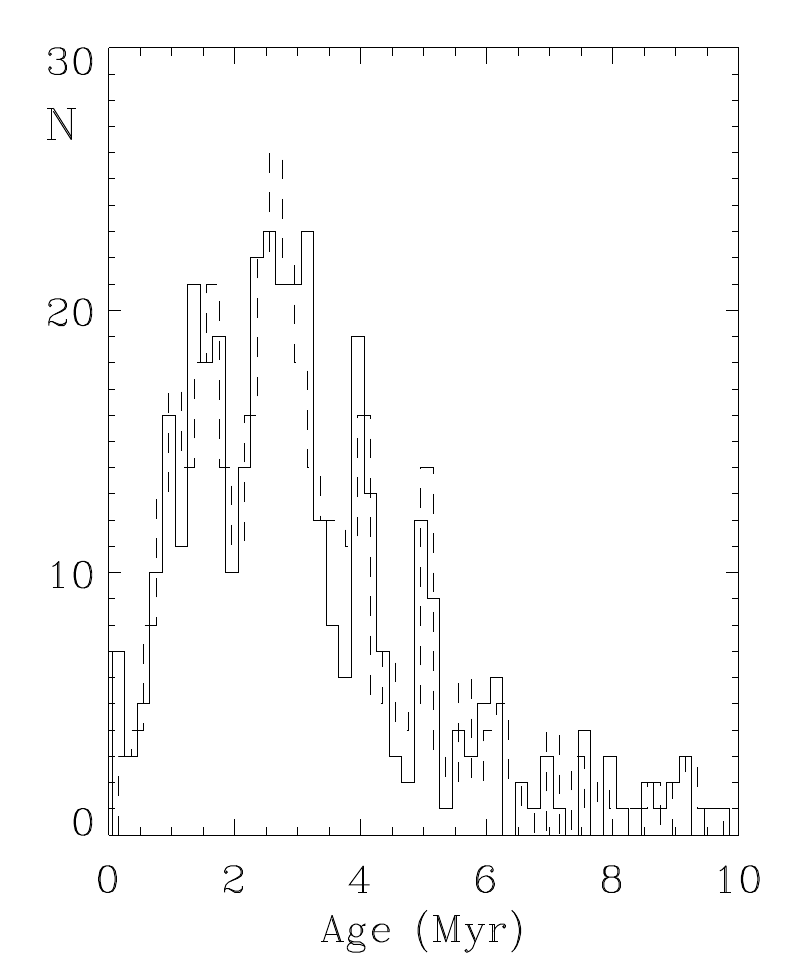}
\caption{Age distribution of PMS members. The median age is 
about 3 Myr with a spread of 6 Myr. This age is rather younger than that estimated 
for the massive stars.}
\label{fig10}
\end{figure}

We have also estimated the age of IC 1848 from low-mass PMS stars 
using the PMS evolution models of \citet{SDF00}. Figure~\ref{fig10} shows the age 
distribution of PMS stars. The majority of PMS stars cover a span of 6 Myr with a 
median age of about 3 Myr. Given the age spread of 3 -- 6 Myr found in other young open 
clusters, such as NGC 2244 \citep{PS02}, NGC 2264 
\citep{PSBK00}, NGC 6530 \citep{SCB00}, and Trumpler 14 and 16 
\citep{HSB12}, the age spread estimated in this work is reasonable, although the age determined for 
PMS stars strongly depends on the PMS evolution models used. 
In Figure~\ref{fig10}, the long tail toward older age may arise from overestimating the age of stars in the 
Kelvin-Helmholtz contraction phase in the PMS models \citep{SBL97,SBC04}, and from stars 
with an edge-on disk lying below the PMS locus \citep{SBCKI08}. It is worth noting that 
some of stars in the long tail may be real, old PMS stars, since the star formation history 
of the W5 region could be quite complex. In order to understand the star formation history 
in W5 West, including IC 1848, the age distribution based on more complete, 
membership selection is needed.

\citet{KM03} estimated a lower limit of 2.4 Myr for the age of IC 1848, assuming a sound speed for the 
ionized gas, and an upper limit of 5 Myr, from the lifetime of O stars. 
\citet{KA11} inferred that the age of the W5 region is younger than 5 Myr using 
the evolution model of \citet{SDF00}. Our estimates are in good agreement 
with those of the previous studies. On the other hand, \citet{Mo72} claimed 
that the age of IC 1848 could not exceed 1 Myr given the existence of the
binary or multiple system, HD 17505, HD 17520, 
and BD+60 586. However, no supporting discussion of the luminosity class 
or the lifetime of such stars was provided in the study, 
and we believe that an age of $\le$ 1 Myr is unlikely. 

\section{THE INITIAL MASS FUNCTION}

As the IMF is a very important tool to understand 
star formation processes, a number of studies have attempted to derive the 
IMF in different star forming environments. \citet{BCM10} reviewed
IMF studies and the relationship with environmental conditions and mass range. Since 
IC 1848 is one of the large-scale star forming regions in the Perseus 
spiral arm, this object must be an ideal laboratory to study the IMF 
in a low metallicity and low spatial density environment. Given the 
fact that the cluster contains a few massive stars it is possible to derive 
the IMF over a wide mass range. 

\begin{figure}
\includegraphics[height=0.52\textwidth]{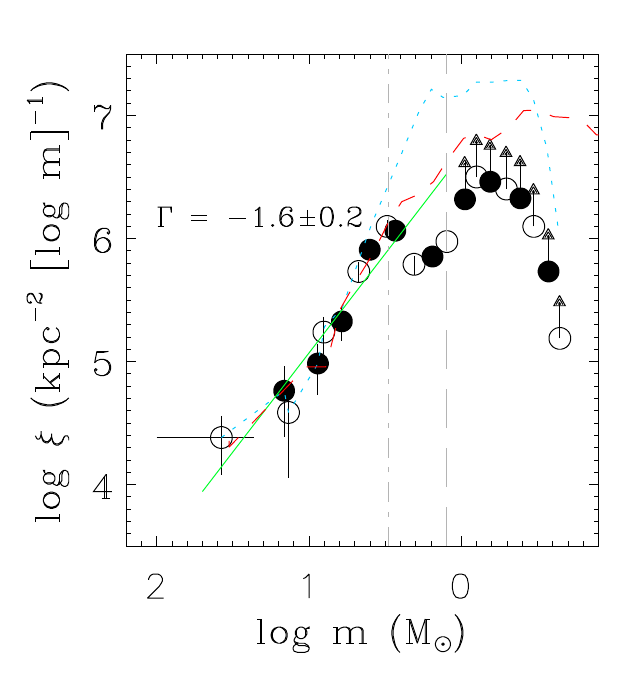}
\caption{The IMF of IC 1848. Circles and the dotted line represent the IMF derived 
from members only and members plus candidates within the pre-main sequence 
locus, respectively. The IMF of NGC 2264 (dashed line -- \citealt{SB10}) is 
overplotted. Dot-dashed and long dashed lines correspond to the completeness 
limits for membership and photometry, respectively. See the main text for details.}
\label{fig11}
\end{figure}

For all post-MS or MS members, the masses were estimated by comparing their 
position in the HRD to post-MS evolutionary tracks. The mass of 
PMS stars, or stars in the PMS locus without any membership 
criterion, were estimated by comparing their position in the HRD to PMS evolution model tracks.
In order to derive the IMF we counted stars within a given mass bin ($\Delta \log m = 0.2$). For the most massive bin, 
the bin size ($\Delta \log m$) was 0.64 because of the small number of massive stars. 
The star counts were then normalized by the logarithmic mass bins and the observed area. The IMF 
of IC 1848 is shown in Figure~\ref{fig11}. To avoid the binning effect we 
shifted the mass bin by 0.1 and re-derived the IMF (open circles), using the same 
procedure as above. 

 In the figure, the circle and dot represent the IMF of members with H$\alpha$ emission 
or YSOs from \citet{KAG08}, while the dotted line corresponds to the IMF including 
all stars within the PMS locus. The shape of the IMF of members in the low-mass regime 
is very similar to that of previous studies for other young open clusters, such as 
NGC 2244 \citep{PS02}, NGC 2264 \citep{PSBK00}, 
NGC 6530 \citep{SCB00}. With complete membership criteria, \citet{SB10} showed 
a smooth increase in the IMF of NGC 2264 down to 0.25 $M_{\sun}$ compared to 
that of \citet{PSBK00}. \citet{SCB00} and \citet{SSB09} have discussed the efficiencies of detecting PMS stars using 
H$\alpha$ photometry (47 \% from a single observation) and MIR excess 
(38 \%), respectively. Hence we attribute the dip in the IMF of IC 1848 
to the incomplete membership selection for the stars in the Kelvin-Helmholtz contraction phase, 
while the incompleteness of our photometry may influence the IMF in the lower mass regime ($< 1.2 M_{\sun}$). 
The IMF of the members ($\geq 1.2 M_{\sun}$) should be a lower 
limit, whereas that of all stars within the PMS locus may be an upper 
limit in the same mass range. 

We first derived the IMF of IC 1848 complete down to $3 M_{\sun}$ through this study. 
The slope ($\Gamma$) of the IMF is about $-1.6 \pm 0.2$ for $\log m > 0.5$. It 
appears to be steeper than the single IMF \citep{Sp55} and the Kroupa IMF \citep{K02}, but 
close to that of \citet{Sc86} for massive stars. The slope of the IMF for IC 1848 is 
also steeper than that of other young open clusters ($-1.3 \pm 0.1$ for NGC 6530 
-- \citealt{SCB00}, $-0.7 \pm 0.1$ for NGC 2244 -- \citealt{PS02}, $-1.2$ for Orion Nebula Cluster (ONC) 
-- \citealt{MLLA02}, $-1.3$ for W5 East -- \citealt{CPO11}, $-1.3 \pm 0.1$ for Trumpler 
14 and 16 -- \citealt{HSB12}, $-1.1 \pm 0.1$ for NGC 6231 -- \citealt{SSB13}). 
On the other hand, the slope ($\Gamma = -1.7 \pm 0.1$) and shape of the IMF 
for NGC 2264 \citep{SB10} are similar to those of IC 1848 in the same mass range. In order to 
arrive at a firm conclusion on the shape of the IMF down to subsolar mass regime, 
further deep optical and/or NIR observation with extensive X-ray observations are needed. 
The complete IMF of the cluster will give an insight into the star formation processes 
in the large-scale star forming region.
 
\begin{figure}
\includegraphics[height=0.45\textwidth]{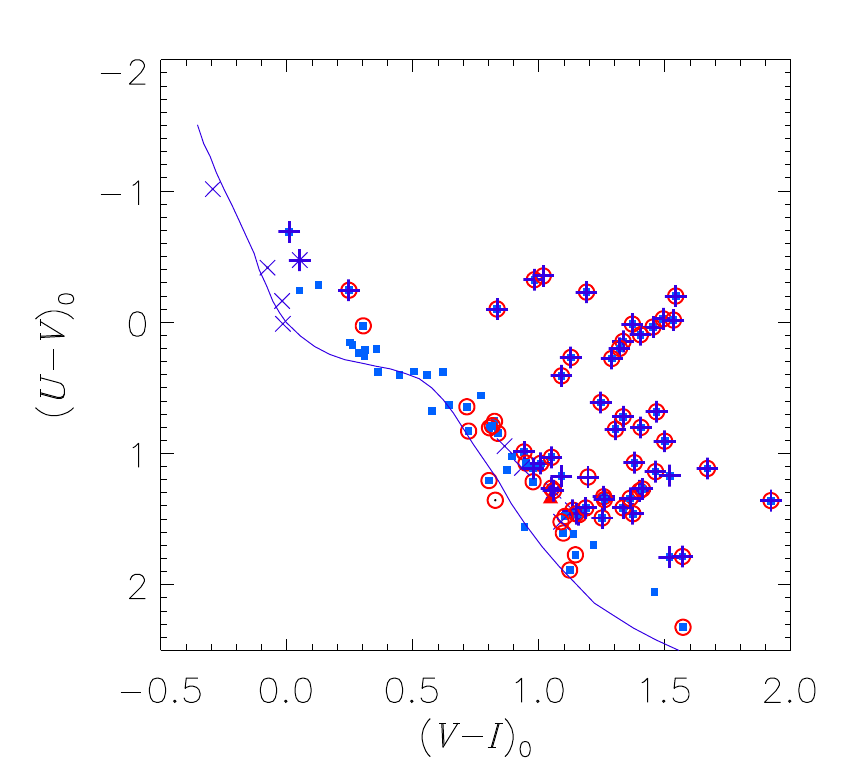}
\caption{The $(U-V)_0$ vs. $(V-I)_0$ diagram. Several pre-main sequence stars 
exhibit a remarkable UV excess driven by their accretion activity. A total of 51 UV excess 
stars (plus symbols) were found . The other symbols are the same as in Figure 2.}
\label{fig12}
\end{figure}

\section{ACCRETION RATES OF PMS STARS WITH A UV EXCESS }

The mass accretion rate gives important information on the disk evolution 
around PMS stars. Since the emergence of the magnetospheric accretion 
model for PMS stars, the physical quantities associated with the accretion process have 
been studied by observation and modeling UV excess emission or/and spectral lines such as 
H$\alpha$, Pa$\beta$, Br$\gamma$, [O I] 6300 \AA, etc, within such a paradigm. 
In this section we estimate the mass accretion rate of those PMS stars with a
UV excess found from our photometric data using the relation between the accretion 
luminosity and $U$-band brightness \citep{GHBC98} and compare the results with 
that of other studies which used independent ways to estimate accretion luminosities 
for different star forming regions.

The first step is to find UV excess stars among PMS members which exhibit 
either H$\alpha$ emission or a MIR excess. Several PMS stars exhibit a UV excess 
in the $[(U-V)_0, (V-I)_0]$ diagram as shown in Figure~\ref{fig12}. The majority 
of stars with a UV excess are late-type members ($V-I \ge 0.7$ mag). These late-type 
members appear to be bluer in $U-V$ than normal MS stars without excess emission at a 
given $(V-I)_0$. The deviation from the expected photospheric color of MS stars 
is larger than the value expected from differential reddening ($\sigma _{E(U-V)} = \sqrt{\sigma ^2 _{E(B-V)} + \sigma ^2 _{E(U-B)} } 
= 0.067$ mag). Hence the large UV excess may be related to the intrinsic properties of late-type 
PMS stars rather than to inappropriate reddening corrections (we have further discussions on the reddening 
corrections for PMS members in Section 3.2). Similar features can 
be found in \citet{RHS00} for the low-mass members in the ONC. The authors 
discussed several sources affecting the calculated UV excess, such as chromospheric 
activity, accretion activity, and the difference between the intrinsic colors of MS and giant stars. 

\citet{RHS00} have examined the influence of chromospheric activity on the size of the 
UV excess using field dMe stars and young stars in the Taurus-Auriga star forming region. 
They suggested that the limit of UV excess from chromospheric activity is about -0.5 mag. 
This value is adopted as a criterion for the PMS stars with a UV excess in the $[(U-V)_0, (V-I)_0]$ 
diagram. A total of 51 members were identified as PMS stars with a UV excess as shown 
in Figure~\ref{fig12} (plus symbols). We computed the $U_{\mathrm{exp}}$ magnitude expected 
for the photospheric color of MS stars, the extinction-corrected $U_{0}$ 
magnitude of stars with UV excess emission, then transformed them 
to luminosity ($L_{U,\mathrm{exp}}$ and $L_{U,0}$) using a bandwidth 
(700 \AA) and zero magnitude flux of $4.22 \times 10^{-9} \mathrm{ergs} \ \mathrm{cm}^{-2}$ \AA$^{-1}$ for 
the Bessell $U$ band \citep{C00}. The accretion luminosity $(L_{\mathrm{acc}})$ was estimated from the 
relation of \citet{GHBC98}: 

\begin{equation}
 \log (L_{\mathrm{acc}}/L_{\sun}) = 1.09 \log (L_{U,\mathrm{exc}}/L_{\sun}) + 0.98
\end{equation}

\noindent where $L_{U,\mathrm{exc}} \equiv L_{U,0} -L_{U,\mathrm{exp}}$. 

\begin{figure*}
\includegraphics[height=0.45\textwidth]{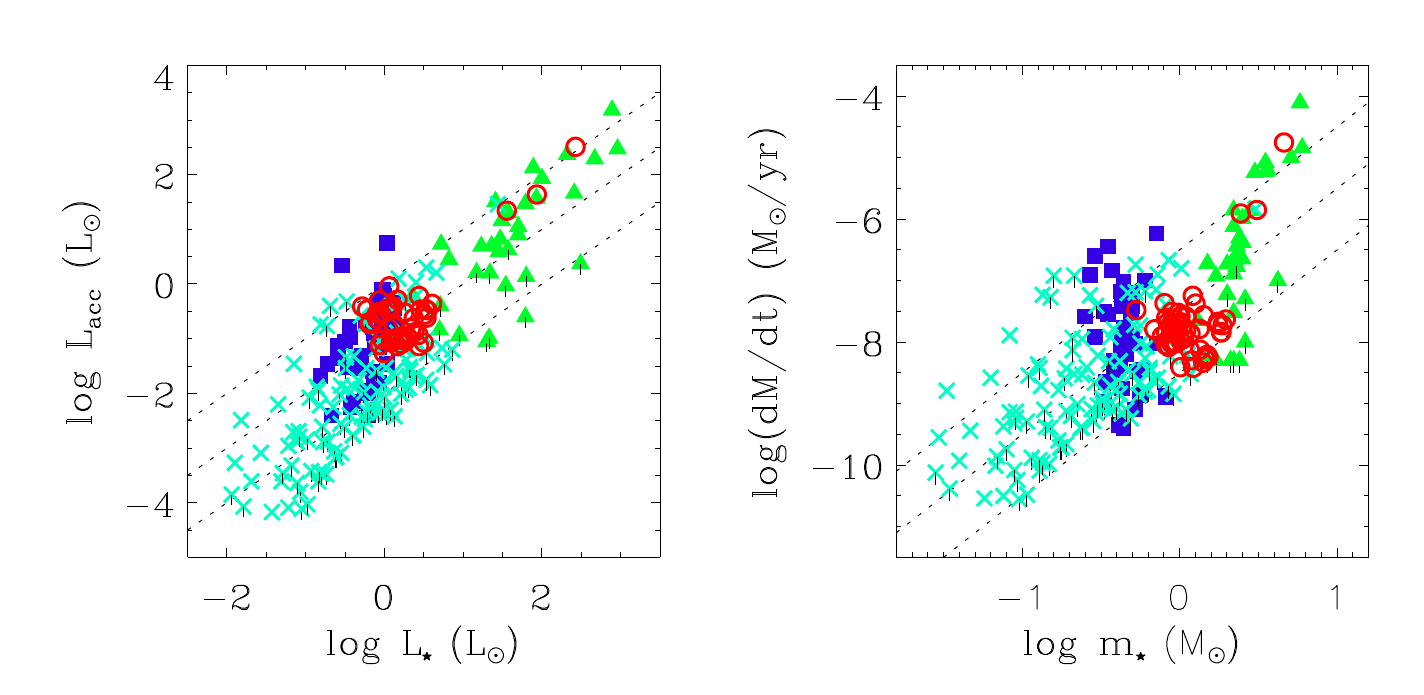}
\caption{Accretion luminosity vs. stellar luminosity (left) and mass accretion 
rate vs. stellar mass (right). Open circles (red) denote our estimates. Squares, 
triangles, and crosses represent the mass accretion rates derived in different star 
forming regions by \citet{HCGD98}, \citet{MCM11}, and \citet{NTR06}, 
respectively.  Dotted lines represent the relations $L_{\mathrm{acc}} \propto L_{\mathrm{stellar}}$ and 
$\dot{M} \propto M^2_{\mathrm{stellar}}$ with an arbitrary constant.}
\label{fig13}
\end{figure*}

In order to obtain mass 
accretion rates we estimated the mass of individual stars from the evolution 
models of \citet{SDF00}. The effective temperature and 
bolometric magnitude of PMS stars obtained above allow us to estimate the radii 
of the stars. With mass ($M_{\mathrm{PMS}}$), radius ($R_{\mathrm{PMS}}$), and accretion 
luminosity ($L_{\mathrm{acc}}$) the mass accretion rate of PMS stars is obtained by 
using the equation below \citep{HCGD98,GHBC98}:

\begin{equation}
\dot{M} = L_{\mathrm{acc}} R_{\mathrm{PMS}} / 0.8 G M_{\mathrm{PMS}} 
\end{equation}
 
\noindent where $G$ and $\dot{M}$ represent the gravitational constant and 
mass accretion rate, respectively. We present the mass accretion rate of 
PMS members with UV excess in Figure~\ref{fig13}. The mass range does not fully 
cover the low-mass stars ($\le 1 M_{\sun}$) as shown in \citet{RHS00} and \citet{MRR12} for the ONC, 
because IC 1848 is more distant than the ONC. In the same mass range our results are 
comparable to those studies by \citet{RHS00} and \citet{MRR12} for the ONC. 
We also plotted results from other studies for different star forming regions 
(\citealt{HCGD98,NTR06,MCM11}, and data therein). The authors used independent 
ways to estimate the accretion luminosity, such as modelled Balmer excess emission 
and the correlation between accretion luminosity and the strength of hydrogen 
recombination lines, Pa$\beta$, and Br$\gamma$. The accretion luminosity and 
mass accretion rate in this study seem to be compatible with the estimates of other 
studies for different star forming regions. 

The mean mass accretion rate of stars in the mass range of $0.5 M_{\sun}$ to 
$2 M_{\sun}$ is about $1.4 \times 10^{-8} M_{\sun} \ \mathrm{yr}^{-1}$. Because of the limited 
mass range of our sample, we cannot confirm whether or not 
$\dot{M}$ is proportional to $M^b_{\mathrm{stellar}}$ ($b = 1.8-2.1$, \citealt{MHC03,MLBHC05,NTR06}). 
However, with the previous estimates for masses below $2 M_{\sun}$ our results are well consistent 
with the previously known correlation between the two quantities. On the other hand, as shown in 
\citet{MCM11}, three intermediate-mass PMS stars exhibit higher mass accretion 
rates than expected from the correlation found for low-mass stars. In that study, 
the authors found that the age of Herbig Ae/Be stars in their sample is systematically 
younger than that of low-mass stars, and therefore the steep slope in the mass accretion rate with 
respect to stellar mass was interpreted as the active accretion of young intermediate-mass PMS 
stars. In this study, we also found that the PMS star (ID 5923) with the highest mass accretion rate is younger than 
the other intermediate-mass PMS stars (2.5-5 $M_{\sun}$). However, \citet{GNTH06} obtained lower mass accretion rates for the same 
Herbig Ae stars. The estimates of \citet{DB11} (except the upper limits), also reveal 3 times lower values for 
Herbig Ae/Be stars. The systematic difference in the mass accretion rate for the Herbig 
Ae/Be stars among authors may have arisen from many uncertain factors 
such as different uses of evolution models, calibration schemes, reddening correction, 
distance, disk geometry with mass, etc. However the details are not within our scope of study based 
on limited data. 

Unlike low-mass star forming regions, our target is 
a young open cluster hosting several high-mass MS stars. Thus, the 
fundamental parameters, such as reddening and distance, can be constrained 
very well to estimate the mass accretion rate of PMS stars ($0.5 \leq M/M_{\sun} 
\leq 5$). Furthermore,  in forthcoming papers as part of the SOS project, homogeneous photometric 
data for many young 
open clusters within 3 kpc from the Sun will provide homogeneous estimates 
of the mass accretion rates and contribute to understanding 
the evolution of the circumstellar disks of PMS stars and formation of planetary systems. 

\section{SUMMARY}
IC 1848 is one of the most interesting targets in the large-scale star forming environment of the
Cas OB6 association. We carried out $UBVI$ and H$\alpha$ photometry for the young 
cluster IC 1848 in the W5 West region as part of the SOS project. This study provided not 
only homogeneous photometric data but also detailed and comprehensive results for IC 1848 
as shown below.

Using the photometric properties of early-type stars, a total of 105 stars were selected 
as members. We identified 397 young stars with MIR excess emission from 
the {\it Spitzer} IRAC data \citep{KAG08} and using our 
H$\alpha$ photometry and spectroscopy by \citet{KA11}, 257 H$\alpha$ emission 
stars were also identified. A total of 462 PMS stars were selected as PMS members.

From the $(U-B, B-V)$ diagram we obtained the reddening of individual 
early-type stars yielding a mean reddening of $\langle E(B-V) \rangle = 0.66 \pm 0.05$ mag. 
In common with most young star clusters or associations, differential reddening was not 
negligible in the observed regions. We constructed the reddening map to correct the 
reddening of PMS members. The photometric data from the
optical to MIR for the early-type stars ($U-B \leq -0.2$) in the observed regions allowed us 
to examine the reddening law of IC 1848. We obtained a consistent ratio of total-to-selective 
extinction of $R_V = 3.1$ over a wide wavelength coverage from the optical to the MIR, and 
concluded that the reddening law toward IC 1848 is normal. 

In order to determine the distance to the cluster, ZAMS fitting to the lower 
boundary of the MS band was carried out both in the reddening-corrected and 
reddening-independent CMDs, and we obtained a distance modulus of $11.7 \pm 0.2$ mag, which is in good agreement 
with that of previous photometric studies. The isochrone fitting to the evolved stars 
gives an age of 5 Myr, while the age distribution of PMS stars shows a median age of 
3 Myr with an age spread of 6 Myr. 

The IMF complete down to $3 M_{\sun}$ was derived from cluster members for the 
first time. Due to the incomplete membership selection of low-mass PMS stars, the slope 
of the IMF ($\Gamma = -1.6 \pm 0.2$) was computed for the limited mass range 
($\log m > 0.5$). That was similar to the Scalo IMF rather than the Salpeter/Kroupa 
IMF. We also found that the IMF of IC 1848 closely resembles that 
of NGC 2264 \citep{SB10} among many young open clusters in the same mass 
range.

We found 51 PMS stars with a strong UV excess from our photometry. The $U$ luminosity 
was transformed to an accretion luminosity by adopting an empirical relation 
from the literature. Finally the mass accretion rate of the PMS stars was estimated by 
using the accretion luminosity, the computed stellar radius, and the stellar mass 
inferred from the PMS evolution models. The stars in the mass range of 
0.5 $M_{\sun}$ to 2 $M_{\sun}$ exhibit a mean mass accretion rate of 
$1.4 \times 10^{-8} M_{\sun} \ \mathrm{yr}^{-1}$. Our result is well consistent with 
the estimates of other studies for different star forming regions. We also found that high-mass 
stars show a higher mass accretion rate than the accretion rate expected from the 
correlation.

\section*{acknowledgments}
The authors express deep thanks to the anonymous referee for
many useful comments and suggestions and Dr. Colette Salyk for useful 
discussions. This work was supported by a National Research Foundation of Korean (NRF) 
grant funded by the Korea Government (MEST) (Grant No.20120005318).


\end{document}